\documentclass[prd,nopacs,showkeys,amsmath,amssymb,superscriptaddress,nofootinbib]{revtex4}

\usepackage{amssymb}
\usepackage{amsmath}
\usepackage{graphicx}
\usepackage{floatflt}

\normalbaselines
\newcommand{\rank}{\mathop{\rm rank}\nolimits}
\newcommand{\sgn}{\mathop{\rm sgn}\nolimits}

\newcommand{\eps}{\varepsilon}
\newcommand{\starr}[2]{{\mathop {\rule{0pt}{0pt}{#1}}\limits^{\,#2}}\rule{0pt}{0pt}}
\renewcommand{\Re}{\mathop{\rm Re}\nolimits}
\renewcommand{\Im}{\mathop{\rm Im}\nolimits}

\begin{document}

\title{Non-minimal Einstein-Maxwell theory: \\ the Fresnel equation and the Petrov classification
of a trace-free susceptibility tensor}
\author{Alexander B. Balakin}
\email{Alexander.Balakin@kpfu.ru} \affiliation{Department of
General Relativity and Gravitation, Institute of Physics, Kazan
Federal University, Kremlevskaya street 18, Kazan 420008, Russia}

\author{Alexei E. Zayats}
\email{Alexei.Zayats@kpfu.ru} \affiliation{Department of General
Relativity and Gravitation, Institute of Physics, Kazan Federal
University, Kremlevskaya street 18, Kazan 420008, Russia}


\begin{abstract}
We construct a classification of dispersion relations for the electromagnetic waves non-minimally coupled to the space-time curvature, based on the analysis of the susceptibility tensor, which appears in the non-minimal Einstein-Maxwell theory.  We classify solutions to the Fresnel equation for the model with a trace-free non-minimal susceptibility tensor according to the Petrov scheme. For all Petrov types we discuss specific features of the dispersion relations and plot the corresponding wave surfaces.
\end{abstract}



\maketitle

\section{Introduction}

The Einstein theory of gravitation includes the postulate that in {\it vacuum} test photons move along null geodesic lines of the corresponding space-time, i.e., the space-time geometry predetermines the photon trajectories. In order to obtain the detailed information about photon behavior in vacuum in the space-time with given symmetry, the classification of the metrics with respect to Lie groups \cite{Exact,PetrovBook} can be used.

The Einstein-Maxwell theory deals with more sophisticated situation, when the propagating electromagnetic waves are coupled to a material {\it medium} \cite{Synge,LL,Maugin,HehlBook}, or to quasi-media of various types (see, e.g., \cite{BBL2012,BD2014,BL2014,BZ2008}). In this case the geometric classification of space-times is not enough, and one needs the additional classification of the electromagnetically active media. What are the objects and the tool of the corresponding classifications? To answer this question we would like to remind two well-known facts.

First, when we are interested in the analysis of space-times from the symmetry point of view, we study the solutions of the equations containing the Lie derivative of the metric along the vector field $\xi^l$, e.g.,  $\pounds_{\xi} g_{ik}=0$ (for the group of isometries), or $\pounds_{\xi} g_{ik}= 2\Phi g_{ik}$ (for the group of conformal motions), etc \cite{Exact,Hallbook}.
In other words, the object of classification is the metric, while a vector field $\xi^i$ (the Killing vector, conformal Killing vector, etc.) is the tool (key element) of classification.

Second, when one deals with the algebraic Petrov classification of the Einstein space-times \cite{PetrovBook,PetrovPaper}, the object of classification is the Weyl tensor, $W^{ikmn}$, the trace-free constituent of the Riemann tensor $R^i_{\ kmn}$ \cite{Exact}, and the tool of classification is the set of eigenbivectors with corresponding eigenvalues. When the space-time is not Einstein one, but is conformally flat, i.e., $W^{ikmn}=0$, an additional classification of the Ricci tensor $R_{ik}$ \cite{Exact,Ricci1,Ricci2} becomes necessary.

The natural question arises: what is the object for media classification in the linear electrodynamics, and what is the appropriate tool? In fact, the answer follows from the works of Tamm \cite{Tamm1,Tamm2}, in which the idea was proposed to use the so-called linear response tensor (or the constitutive tensor in the alternative terminology) $C^{ikmn}$,  which links the excitation tensor, $H^{ik}$ with the Maxwell tensor $F_{mn}$, by the linear relationship $H^{ik}=C^{ikmn}F_{mn}$. This tensor $C^{ikmn}$ depends on the properties of the medium only, contains the information about the dielectric permittivity, magnetic permeability, magneto-electric cross effects \cite{LL,Maugin,HehlBook}, and thus can be the intrinsic characteristic of the electromagnetically active medium. We follow this approach, and consider below the tensor $C^{ikmn}$ as the object for classification; as for the corresponding tool, the situation in general case depends on the used approach.

Historically, the first attempt to classify the linear response tensor $C^{ikmn}$ is connected with the formalism of optical metrics (see, e.g., \cite{Gordon,PMQ,Ehlers} and \cite{Synge,PerlickBook}). According to the idea of Gordon, one can introduce the metric of some effective space-time, in which the photons, coupled, in fact, to the medium in the real space-time, ``propagate'' along the null geodesic lines attributed to this effective space-time. It was proved, in particular, that in terms of the optical metric, $\frak{g}^{ik} = n^{-2}\left[g^{ik} + (n^2-1)U^iU^k\right]$, found for the spatially isotropic medium with the refraction index $n$ and magnetic permeability $\mu$, moving with the macroscopic velocity $U^i$, the linear response tensor  $C^{ikmn}_{({\rm isotr})} = \frac{n^4}{2\mu} \left[\frak{g}^{im}\frak{g}^{kn}-\frak{g}^{in}\frak{g}^{km}\right]$ has the same form, as the one for the vacuum $C^{ikmn}_{({\rm vac})}=\frac12 (g^{im}g^{kn}-g^{in}g^{km})$ (see, e.g., \cite{PMQ}). Based on this idea, in \cite{BalZim2005} the tensor of linear response was decomposed using two associated metrics $\frak{g}^{ik}_{A}$ and $\frak{g}^{ik}_{B}$ for the case of uniaxial medium. The decomposition of the same type, but for the colored linear response tensor $C^{ikmn}_{({\rm a})({\rm b})}$ with respect to color (effective) metrics $\frak{g}^{ik (\alpha)}_{({\rm a})}$, was made in \cite{BDZ2007,BDZ2008} in the framework of the $SU(N)$ symmetric Einstein-Yang-Mills theory ($({\rm a})$ is the group index). As the result, it was shown that the classification of the linear response tensor with respect to effective (associated, color, optical) metrics is very useful for the spatially isotropic and uniaxial media, but this method is not effective, when the medium has biaxial symmetry \cite{PerlickBook,BalZim2005}. In other words, the effective metrics can be used as a tool for the classification of the linear response tensor, but the classification is not complete.

Generally, the linear response tensor is skew-symmetric with respect to permutation of indices inside the pairs $[ik]$ and $[mn]$; the symmetry of the pairs is not obligatory.
When  $C^{ikmn} \neq C^{mnik}$ the decomposition of this constitutive tensor with respect to irreducible tensor elements shows that in addition to two standard symmetric permittivity tensors, $\eps^{im}$, ${(\mu^{-1})}_{pq}$ and one non-symmetric cross-effect (pseudo-)tensor $\nu^{im}$ \cite{Maugin}, the so-called skewonic and axionic parts appear (see, e.g., \cite{HehlBook,Hehl2,Hehl3,Hehl4,Itin2013,WTN,Itin2015,Itin2016} and \cite{Ni1977,CFJ1990,Itin2004,Hehl1,Itin2007,Itin2008,Hehl2008} for details and references). The so-called premetric axiomatic scheme in the electrodynamics elaborated by Hehl and colleagues (see, e.g., \cite{Hehl5,Hehl6,Hehl7,Itin2009}) can be also considered as a realization of the idea of the representation of the constitutive tensor $C^{ikmn}$ with respect to metric (but now there is no basic metric, and the metric, which the authors extract from the constitutive tensor, is not, generally speaking, an analog of Gordon's optical metric).

The non-minimal Einstein-Maxwell theory (see, e.g., \cite{NM1,NM2,NM3,NM4,NM5,NM6} for details and references) can also be described in terms of (quasi-)vacuum electrodynamics. Indeed, the coupling of photons to the space-time curvature produces variations of the phase and group velocities of the electromagnetic waves propagating in the non-minimal vacuum (see, e.g., \cite{NM2,NM4,BBL2012,BD2014,BZ2008}). Non-minimal coupling induces the effect of birefringence \cite{NM7}, the Cherenkov effect \cite{NM8}, the optical activity \cite{NM9}, i.e., in the framework of the non-minimal Einstein-Maxwell theory the physical vacuum behaves as a specific medium, the quasi-medium. The main purpose of this paper is to classify the non-minimally extended linear response tensor $C^{ikmn}$ and related Fresnel equations (the objects of classification) in terms of the Petrov classification. The non-minimal Einstein-Maxwell theory is the unique theory, for which the Petrov scheme can be applied for both classifications: of the space-time and of the non-minimal electromagnetically active quasi-medium.

The paper is organized as follows. In Subsect.~\ref{IIsectA} we recall the key elements of the Einstein-Maxwell theory. In Subsect.~\ref{IIsectB} we describe the geometrical optics approach, the Fresnel equations for a general linear response tensor $C^{ikmn}$ and their algebraic structure. In Sect.~\ref{IIIsect} we discuss the essential details of the non-minimal theory and the properties of the non-minimal susceptibility tensor. In Sect.~\ref{IV} we analyze the structure of the Fresnel equation for all classes appeared in the Petrov classification of the space-times and plot the corresponding wave surfaces in Sect.~\ref{V}. Results and outlook are formulated in Sect.~\ref{VI}.

\section{Einstein-Maxwell theory and the problem of electromagnetic wave propagation}\label{IIsect}

\subsection{Basic formalism of the medium electrodynamics}\label{IIsectA}

The Einstein-Maxwell theory deals with the action functional
\begin{equation}
S_{({\rm EM})} = \int d^{4} x \sqrt{-g}\ \left(
\frac{R}{\kappa}+\frac{1}{2}\, C^{ikmn} F_{ik}
F_{mn} \right)
\,,
\label{act0}
\end{equation}
where $R$ is the Ricci scalar. The quantity $F_{ik} \equiv \nabla_i A_k-\nabla_k A_i$ is the Maxwell tensor, defined on the base of the electromagnetic field potential four-vector $A_i$. The information concerning specific features of interactions in the electromagnetically active medium (or quasi-medium) is encoded in the linear response tensor $C^{ikmn}$ (it can be called also as a constitutive tensor). Due to the structure of the second term in (\ref{act0}) this tensor possesses evident symmetry of indices
\begin{equation}
C^{ikmn} = -C^{kimn} = C^{mnik} =-C^{iknm}. \label{e1}
\end{equation}
Variation of the action functional (\ref{act0}) with respect to potential
$A_i$ yields the electrodynamic equations
\begin{equation}
\nabla_k \left(C^{ikmn}F_{mn}\right)=0\,. \label{e01}
\end{equation}
The equations $\nabla_k \starr{F}{*}^{ik}=0$ is the consequence of the Maxwell tensor definition. Hereafter, the asterisk denotes the dualization procedure, e.g.,
$\starr{F}{*}^{ik}\equiv \frac12 \epsilon^{ikmn}F_{mn}$ for the second-rank tensor, where $\epsilon^{ikmn}$ is the completely skew-symmetric Levi-Civita tensor. If we deal with the fourth-rank tensors, the dualization can be one of two types: left dualization, ${\mathstrut}^*\!{C}_{ikmn}\equiv \frac12 \epsilon_{ikls}{C^{ls}}_{mn}$, and the right one, ${C\mathstrut}^*_{ikmn}\equiv \frac12 \epsilon_{mnls}{C_{ik}}^{ls}$. When the left-dual tensor and the right-dual tensor coincide, we will use the centered asterisk, $\starr{C}{*}_{ikmn}$. The double-dual tensor, for which both types of dualization are applied simultaneously, we will denote as $\starr{C}{**}_{ikmn}$, i.e., $\starr{C}{**}_{ikmn} \equiv \frac14 \epsilon_{ikls}\epsilon_{mnpq} C^{lspq}$.

The tensor $C^{ikmn}$ contains information about dielectric and magnetic permeabilities
as well as about the magneto-electric coefficients \cite{LL,Maugin,HehlBook}. Using the medium velocity four-vector $U^i$, normalized such that $U_iU^i = 1$, one can decompose $C^{ikmn}$ uniquely as follows
\begin{align}\label{Cem}
    C^{ikmn}&=\frac{1}{2}\left(\eps^{im}U^kU^n+\eps^{kn}U^iU^m-\eps^{in}U^kU^m-\eps^{km}U^iU^n\right)-{}\nonumber\\
    {}&-\frac{1}{2}(\mu^{-1})_{ls}\epsilon^{iklp}\epsilon^{mnsq}U_pU_q
    -\frac{1}{2}\left[{\epsilon^{ik}}_{lp}(U^mU^p\nu^{ln}-U^nU^p\nu^{lm})+
    {\epsilon^{mn}}_{lp}(U^iU^p\nu^{lk}-U^kU^p\nu^{li})\right].
\end{align}
Here, $\eps_{im}$ is the dielectric permittivity tensor, $\mu^{-1}_{pq}$ is the magnetic impermeability tensor and $\nu_{lm}$ is the magneto-electric coefficients pseudo-tensor. These quantities are defined through
\begin{align}\label{definition}
    \eps_{im}&=2\,C_{ikmn}U^kU^n, \\
    (\mu^{-1})_{ls}&=-2\starr{C}{**}_{lksn}U^kU^n, \\
    \nu_{lm}&=2\,{\mathstrut}^*\!{C}_{lsmn}U^sU^n,
\end{align}
They are space-like, i.e., they are orthogonal to $U^i$ with respect to each of their indices. The symmetry conditions (\ref{e1}) for the tensor $C_{ikmn}$ yield
\begin{gather}
\eps_{im}=\eps_{mi},\ \mu^{-1}_{ls}=\mu^{-1}_{sl}
\end{gather}
and indicates that our model has no skewons. The tensor $\nu_{lm}$ is generally non-symmetric.
If, in addition, the linear response tensor satisfies the relation
\begin{equation}
C_{ikmn}+C_{imnk}+C_{inkm}=0, \label{e2}
\end{equation}
the model is free from axions and the trace of the magneto-electric coefficients tensor has to vanish ${\nu_m}^m=0$. The last condition is also known as the Post constraint (see \cite{Postbook,Hehl1}).

In the vacuum the linear response tensor has the simplest form
\begin{equation}
C^{ikmn}_{({\rm vac})}=\frac{1}{2}(g^{im}g^{kn}-g^{in}g^{km})\,.
\label{evac}
\end{equation}
The difference ${\chi}^{ikmn}\equiv  C^{ikmn} - \frac{1}{2}(g^{im}g^{kn}-g^{in}g^{km})$
is called the susceptibility tensor.

\subsection{Fresnel equation}\label{IIsectB}

When the short-wavelength electromagnetic radiation propagate in the curved space-time, the approximation called
the geometrical optics is the appropriate tool for analysis. In this approach the potential four-vector and the field strength tensor can be represented, respectively, as
\begin{equation}
A_m=a_me^{i\Theta}\,,\quad F_{mn} = i (K_ma_n-K_na_m)e^{i\Theta}\,,
\label{Petrov58}
\end{equation}
where $\Theta$ is the phase, $a_m$ is a slowly varying
amplitude, and $K_m$ is a wave four-vector defined as the gradient of the phase, $K_m=\nabla_m \Theta$.
In the leading-order approximation, the Maxwell equations can be reduced to the system of algebraic equations
\begin{equation}
C_{pqrs}K^qK^ra^s = 0\,.
\label{main}
\end{equation}
This set of linear equations with respect to $a_s$ is evidently admits the pure gauge
solution $a_s\sim K_s$. In order to find non-trivial solutions to Eq.~(\ref{main}) we have to require
that the rank of the matrix $\mathfrak{A}_{ij}\equiv C_{ipqj}K^pK^q$ is less than three, i.e., all third-order subdeterminants of $\mathfrak{A}_{ij}$ have to vanish
\begin{equation}
\Delta^{ij}\equiv\frac{1}{3!}\epsilon^{ipmn}\epsilon^{jqrs}\mathfrak{A}_{pq}\mathfrak{A}_{mr}\mathfrak{A}_{ns}=0.
\end{equation}
After some routine calculations we can rearrange the expression for $\Delta^{ij}$ in the form
\begin{gather}
  \Delta^{ij}=\frac18 K^iK^j\cdot {\cal G}^{pqrs}K_pK_qK_rK_s, \label{Id1}\\
  {\cal G}^{pqrs}\equiv -\frac{4}{3}\,C^{ipmq}C^{krns} \starr{C}{**}_{ikmn}. \label{TRtensor}
\end{gather}
Thus, Eq.~(\ref{main}) admits non-trivial solutions when the four components of $K_p$ satisfy the
Fresnel equation, which is usually called the dispersion relation:
\begin{equation}
T[K]\equiv{\cal G}^{pqrs}K_pK_qK_rK_s=0.\label{FrenEq}
\end{equation}
The tensor ${\cal G}^{pqrs}$ in the form (\ref{TRtensor}) is known as the Kummer tensor (see, e.g. \cite{Hudson,Zund,Favaro2}),
while its totally symmetric part ${\cal G}^{(pqrs)}$ is usually called the Tamm-Rubilar tensor (see, e.g., \cite{Tamm2,Rubilar2002,Itin2009,Schuler2010}). It is worth noting that, first, the factor $-4/3$ in (\ref{TRtensor}) is chosen merely for the sake of convenience: in the vacuum case, the Tamm-Rubilar tensor and the dispersion relation take the simplest form with a unit factor
\begin{equation}
{\cal G}^{(pqrs)}_{(\rm{vac})}=g^{(pq}g^{rs)}, \quad T_{(\rm{vac})}[K]={\cal G}^{pqrs}_{(\rm{vac})}K_pK_qK_rK_s=(K_pK^p)^2=0;
\end{equation}
second, the Fresnel equation (\ref{FrenEq}) actually does not change, if the tensor $C_{ikmn}$ transforms as follows
\begin{equation}\label{conformtrans}
  C_{ikmn} \rightarrow \Omega\cdot C_{ikmn}, \quad \Omega\neq 0,
\end{equation}
because the Tamm-Rubilar tensor and Eq.~(\ref{FrenEq}) are homogeneous with respect to the tensor $C_{ikmn}$ components.

On the other hand, the Fresnel equation (\ref{FrenEq}) is a quartic homogeneous equation in the wave vector components and it gives $K_p$ up to a factor. Therefore Eq.~(\ref{FrenEq}) defines a quartic surface in a three-dimensional projective space $\mathbb{R}P^3$. This surface may possess, in principle, isolated singular points (maximal number of such points is sixteen \cite{Hudson}) and/or singularities located on a line.  Positions of these singularities are determined by the condition
\begin{equation}\label{singEq}
  \frac{\partial T[K]}{\partial K_s}=4{\cal G}^{(pqrs)}K_pK_qK_r=0.
\end{equation}
To clarify the physical significance of Eq.~(\ref{singEq}), we should remind that the system (\ref{main}) with the wave vector $K_p$ satisfying the dispersion relation (\ref{FrenEq}) gives only one non-trivial polarization vector $a_s$, when $\rank (\mathfrak{A}_{ij})=2$. If $\rank (\mathfrak{A}_{ij})=1$ we have two non-trivial polarizations for the fixed wave vector and, as a corollary, we have no birefringence phenomenon along this direction in our (quasi-)medium. Due to differential consequence of the identity~(\ref{Id1})
$$\frac{1}{2!}\epsilon^{ipmn}\epsilon^{jqrl}\mathfrak{A}_{pq}\mathfrak{A}_{mr}\frac{\partial\mathfrak{A}_{nl}}{\partial K_s}=\frac{1}{8}K^iK^j\frac{\partial T[K]}{\partial K_s}+\frac{1}{8}T[K]\left(g^{is}K^j+g^{js}K^i\right),$$
we can conclude that the necessary condition for existence of such directions is Eq.~(\ref{singEq}).

\section{Non-minimal Einstein-Maxwell model}\label{IIIsect}

\subsection{Non-minimal susceptibility tensor}

When one speaks about the three-parameter non-minimal Einstein-Maxwell model, one deals with the specific action functional
\begin{equation}
S_{({\rm NMEM})} = \int d^{4} x \sqrt{-g}\ \left(
\frac{R}{\kappa}+\frac{1}{2}\,F_{ik}F^{ik}+\frac{1}{2}\, {\cal R}^{ikmn} F_{ik}
F_{mn} \right),
\label{act1}
\end{equation}
where tensor ${\cal R}^{ikmn}$ is defined as follows
\begin{eqnarray}
{\cal R}^{ikmn}\equiv
\frac{q_1}{2}R\,(g^{im}g^{kn}-g^{in}g^{km}) + \frac{q_2}{2}(R^{im}g^{kn} - R^{in}g^{km} + R^{kn}g^{im}
-R^{km}g^{in}) + q_3 R^{ikmn}.
\end{eqnarray}
It is composed of the Ricci scalar $R$, the Ricci tensor $R_{ik}$, and the Riemann curvature tensor $R^{ikmn}$ (see, e.g., \cite{NM6} for references and terminology); the phenomenological parameters $q_1$, $q_2$, $q_3$ describe the non-minimal coupling of electromagnetic and gravitational fields. Obviously, the action of the non-minimal Einstein-Maxwell model looks like (\ref{act0}), where the linear response tensor takes the form
\begin{equation}
C^{ikmn}= \frac{1}{2}\left(g^{im}g^{kn}-g^{in}g^{km}\right)+ {\cal R}^{ikmn}\,.
\label{electro2}
\end{equation}
The latter means that the tensor ${\cal R}^{ikmn}$ plays the role of the non-minimal susceptibility tensor.

Let us discuss some properties of ${\cal R}^{ikmn}$ and $C^{ikmn}$. Firstly, they inherit symmetries of indices from the curvature tensor,
\begin{gather}
{\cal R}^{ikmn} = -{\cal R}^{kimn} = {\cal R}^{mnik} =-{\cal R}^{iknm}, \\
{C}^{ikmn} = -{C}^{kimn} = {C}^{mnik} =-{C}^{iknm}
\label{electro3}
\end{gather}
and the Bianchi-type identity
\begin{gather}
{\cal R}_{ikmn} + {\cal R}_{inkm} +  {\cal R}_{imnk} =0,\\
{C}_{ikmn} + {C}_{inkm} +  {C}_{imnk} =0.
\label{electro4}
\end{gather}
Thus, the gravitational field non-minimally coupled with curvature can be considered as a quasi-medium
without ``skewons'' and ``axions''.

Secondly, the non-minimal susceptibility tensor ${\cal R}^{ikmn}$ can be rewritten in terms
of irreducible parts of the curvature tensor (see \cite{Exact})
\begin{equation}\label{decomposition}
  {\cal R}^{ikmn}=\zeta_1G^{ikmn}+\zeta_2E^{ikmn}+\zeta_3W^{ikmn},
\end{equation}
where $W^{ikmn}$ is the traceless Weyl tensor,
\begin{equation}
  W^{ikmn}=R^{ikmn}-\frac12 \left(R^{im}g^{kn}+R^{kn}g^{im}-R^{in}g^{km}-R^{km}g^{in}\right)+\frac{R}{6}\left(g^{im}g^{kn}-g^{in}g^{km}\right),
\end{equation}
and
\begin{gather}
  E^{ikmn}=\frac12 \left(S^{im}g^{kn}+S^{kn}g^{im}-S^{in}g^{km}-S^{km}g^{in}\right),\quad S^{mn}=R^{mn}-\frac14 R g^{mn},\quad S_m^m=0,\\
  G^{ikmn}=\frac{R}{12}\left(g^{im}g^{kn}-g^{in}g^{km}\right).
\end{gather}
The parameters $\zeta_1$, $\zeta_2$, and $\zeta_3$ are connected with $q_1$, $q_2$, and $q_3$ by the linear relations
\begin{equation}
  \zeta_1=6q_1+3q_2+q_3,\quad \zeta_2=q_2+q_3,\quad \zeta_3=q_3.
\end{equation}
It is worth noting that the tensor $E^{ikmn}$ does not change after double duality procedure, while $W^{ikmn}$ and $G^{ikmn}$ change their sign
\begin{equation}\label{duality}
  \starr{W}{**}_{ikmn}=-W_{ikmn},\quad \starr{G}{**}_{ikmn}=-G_{ikmn},\quad \starr{E}{**}_{ikmn}=E_{ikmn}.
\end{equation}

\subsection{Trace-free susceptibility tensor model}

In this paper, we focus on the model with a special type of the non-minimal susceptibility tensor ${\cal R}_{ikmn}$:
we will consider the case, when
\begin{equation}\label{Rdual}
  \starr{\cal R}{**}_{ikmn}=-{\cal R}_{ikmn}
\end{equation}
and, as a consequence,
\begin{equation}\label{Cdual0}
  \starr{C}{**}_{ikmn}=-{C}_{ikmn}.
\end{equation}
This relation means that the decomposition (\ref{decomposition}) does not contain the tensor $E_{ikmn}$, or, in other words, we have to require $\zeta_2 E_{ikmn}=0$.  It is possible in two cases. \\
{\noindent {\it (i)}}  The first case realizes when $\zeta_2=q_2+q_3=0$; in this case the tensor $E_{ikmn}$ and thus the Ricci tensor $R_{im}$ can be arbitrary and the second term in (\ref{decomposition}) is provided to be vanishing due to the choice of the phenomenological parameters $q_2$ and $q_3$. \\
{\noindent {\it (ii)}} In the second case, we assume that $E_{ikmn}=0$; it means the traceless part of the Ricci tensor, $S_{im}$, vanishes. This variant corresponds to all Einstein space-times, $R_{im}=-\Lambda g_{im}$, $R=-4\Lambda$ (Schwarzschild, Kerr, de Sitter space-times etc.). Thus, when we consider light propagation on a certain Einstein space-time background in framework of the non-minimal theory, we deal with our model. \\
Below, in the analysis of Fresnel equation, we do not attract the attention of Readers to the question: which version we use, the first or the second one; for both cases the algebraic structures of the susceptibility tensor coincide and the analysis is identical.


The linear response tensor $C^{ikmn}$ can be written as follows
\begin{equation}\label{Cdual}
  C^{ikmn}=\frac12 \left[1+\frac{6q_1+3q_2+q_3}{6}\,R\right]\left(g^{im}g^{kn}-g^{in}g^{km}\right)+q_3W^{ikmn}.
\end{equation}
Since this tensor satisfies the Bianchi identity (\ref{e2}) and the duality constraint (\ref{Cdual0}) $\starr{C}{**}_{ikmn}=-{C}_{ikmn}$,
the effective tensors of dielectric permittivity and magnetic impermeability coincide with each other
\begin{equation}\label{efftensors}
  \eps_{im}=\mu^{-1}_{im}.
\end{equation}
The magneto-electric coefficients tensor in this case becomes symmetric with respect to its indices and traceless
\begin{equation}\label{efftensors2}
  \nu_{im}=\nu_{mi}, \quad \nu_m^m=0.
\end{equation}
The first term in (\ref{Cdual}) describes an isotropic part of the tensor $C^{ikmn}$ related to the traces of $\eps_{im}$ and $\mu^{-1}_{im}$
\begin{equation}\label{efftrace}
  \eps_{m}^m={\mu^{-1}}_m^m= 3\eps,\quad \eps\equiv 3 \left[1+\frac{6q_1+3q_2+q_3}{6}\,R\right].
\end{equation}
When $\eps\neq0$, the expression (\ref{Cdual}) can be rearranged as follows
\begin{equation}\label{CdualX}
  C^{ikmn}=\eps\left[\frac12\left(g^{im}g^{kn}-g^{in}g^{km}\right)+\frac{q_3}{\eps}\,W^{ikmn}\right],
\end{equation}
and, due to homogeneity of the Fresnel equation (\ref{FrenEq}), we may drop the factor in front of brackets in Eq.(\ref{CdualX}). As a result, the effective susceptibility tensor $\chi_{ikmn}$ appears to be proportional to the Weyl tensor, $\chi_{ikmn}=(q_3/\eps)W_{ikmn}$, and therefore our model can be indicated as the trace-free susceptibility tensor model.

At last, it is worth mentioning that for the considered model the linear response tensor has 11 independent components --- 6 components of the dielectricity tensor $\eps_{im}$ and 5 components of the magneto-electric tensor $\nu_{im}$. On the other hand, this number can be calculated by another way --- the Weyl tensor has 10 independent components, the eleventh one is the factor in front of $\frac12 (g^{im}g^{kn}-g^{in}g^{km})$ in (\ref{Cdual}).

The Kummer tensor ${\cal G}^{pqrs}$, in accordance with (\ref{Cdual0}), can be rewritten as
\begin{equation}
  {\cal G}^{pqrs}=\frac43 C^{ipmq}C^{krns}\,C_{ikmn}.
\end{equation}
Substituting here the representation for the linear response tensor $C^{ikmn}$ (\ref{Cdual}), we obtain
\begin{equation}
  {\cal G}^{pqrs}=\eps^3\, g^{pq}g^{rs}-2\eps\, q_3^2\, W^{ipmq}W_{i\ m}^{\ r\ s}+\frac43q_3^3 \, W^{ipmq}W^{krns}W_{ikmn}+\hbox{non-sym. terms},
\end{equation}
and, after symmetrization with respect to indices
\begin{align}
  {\cal G}^{(pqrs)}&=\left[\eps^3-\frac18 \eps\, q_3^2\, W_{ikmn}W^{ikmn} +\frac{1}{24}q_3^3 \, W_{ijkl}W^{klmn}{W^{ij}}_{mn}\right]g^{(pq}g^{rs)}-4\eps\, q_3^2 B^{pqrs}+{}\nonumber \\
  &{}+\frac43 q_3^3\,\left(B^{pqlm}W^{r\ s}_{\ l\ m}+B^{prlm}W^{s\ q}_{\ l\ m}+B^{pslm}W^{q\ r}_{\ l\ m}\right),\label{Grepr}
\end{align}
where $B^{pqrs}$ is the totally symmetric traceless tensor of the second order with respect to $W^{ikmn}$, also known as the Bel-Robinson tensor \cite{PenroseRindler},
\begin{equation}
B^{pqrs}=\frac14 \left(W^{pkqm}W_{\ k\ m}^{r\ s}+\starr{W}{*}^{pkqm}\,\starr{W}{*}_{\ k\ m}^{r\ s}\right).
\end{equation}
Thus, if we need a classification of the dispersion relations for the trace-free susceptibility tensor model, we should evidently apply the classification of the Weyl tensor.

\subsection{Our further strategy}

In the framework of the trace-free susceptibility tensor model, the non-minimal susceptibility tensor ${\cal R}^{ikmn}$, and thus the linear response tensor $C^{ikmn}$ and the Tamm-Rubilar tensor ${\cal G}^{(pqrs)}$ can be represented in terms of eleven quantities: ten components of  the Weyl tensor, and one real scalar $\eps$ related to the isotropic part of the tensor $C^{ikmn}$ (see (\ref{efftrace})). Then using the classification of the Weyl tensor given by Petrov \cite{PetrovBook}, one can classify the Tamm-Rubilar tensor and therefore describe all the types of electromagnetic waves coupled to curvature for our model.

In the geometrical point of view, this scheme is firmly associated with the type classification of quartic surfaces in the three-dimensional projective space $\mathbb{R}P^3$, which are defined by the Fresnel equation (\ref{FrenEq}) with (\ref{Grepr}). The main purpose of such classification is to find and investigate singularities of these surfaces, their number and properties. In order to illustrate each surface type, we will depict a set of corresponding wave surfaces.

In this paper we make just the first step towards a complete classification of the dispersion relations. To achieve it we have to consider, in principle, a combined classification of the Weyl and Ricci tensors, but it is idea for next few year.

\section{Petrov classification of the model with trace-free non-minimal susceptibility tensor}\label{IV}

\subsection{The tools for analysis: the Newman-Penrose formalism and Petrov classification}

It is well-known that the Weyl tensor can be classified according to the Petrov scheme \cite{Exact,PetrovBook}, thus providing the corresponding classification of the trace-free susceptibility tensor.

We follow the standard Newman-Penrose formalism, but in order to avoid differences, we fix the definitions given in the book \cite{Ch}. The signature of the metric is $\{+---\}$, and the null tetrads
\begin{equation}
e^p_{(1)}=l^p\,,\quad e^p_{(2)}=n^p\,,\quad e^p_{(3)}=m^p\,,\quad e^p_{(4)}=\bar{m}^p,
\label{Petrov1}
\end{equation}
where $l^p$ and $n^p$ are real vectors, $m^p$ and $\bar{m}^p$ compose the complex-conjugate pair of vectors,
satisfy the conditions
\begin{gather}
l^pl_p=n^pn_p=m^pm_p=\bar{m}^p\bar{m}_p=0,
\quad l^pm_p=l^p\bar{m}_p=n^pm_p=n^p\bar{m}_p=0,\nonumber \\
l^pn_p=1,\quad m^p\bar{m}_p=-1.
\label{Petrov2}
\end{gather}
The corresponding tetrad components of a vector $F_p$ are denoted as
\begin{equation}\label{tetrad}
  F_{(1)}=F_pl^p,\quad F_{(2)}=F_pn^p,\quad F_{(3)}=F_pm^p,\quad F_{(4)}=F_p\bar{m}^p.
\end{equation}
When the vector $F_p$ are real one, we obtain that $F_{(4)}=\bar{F}_{(3)}$.

In these terms, the space-time metric $g^{ik}$ and the basic selfdual, $U^{ik}$, $V^{ik}$, $M^{ik}$, and anti-selfdual bivectors, $\bar{U}^{ik}$, $\bar{V}^{ik}$, $\bar{M}^{ik}$, can be written as follows
\begin{equation}
g^{pq}=l^pn^q+l^qn^p-m^p\bar{m}^q-m^q\bar{m}^p \,,
\label{Petrov3}
\end{equation}
\begin{gather}
U^{ik}=-2n^{[i}\bar{m}^{k]}, \quad
V^{ik}=2l^{[i}m^{k]}, \quad
M^{ik}=2m^{[i}\bar{m}^{k]}-2l^{[i}n^{k]},\\
\bar{U}^{ik}=-2n^{[i}{m}^{k]}, \quad
\bar{V}^{ik}=2l^{[i}\bar{m}^{k]}, \quad
\bar{M}^{ik}=-2m^{[i}\bar{m}^{k]}-2l^{[i}n^{k]}.
\label{Petrov33}
\end{gather}
Using them, we can reconstruct the tensor $q_3W^{ikmn}$:
\begin{gather}
q_3 W^{ikmn}= -\Psi_0U^{ik}U^{mn} -\Psi_1(U^{ik}M^{mn}+M^{ik}U^{mn})-{}\nonumber\\
{}-\Psi_2(V^{ik}U^{mn}+U^{ik}V^{mn}+M^{ik}M^{mn}) -\Psi_3(V^{ik}M^{mn}+M^{ik}V^{mn})
-\Psi_4V^{ik}V^{mn} -{}\nonumber\\
{}-\bar{\Psi}_0\bar{U}^{ik}\bar{U}^{mn} -\bar{\Psi}_1(\bar{U}^{ik}\bar{M}^{mn}+\bar{M}^{ik}\bar{U}^{mn})-\bar{\Psi}_2(\bar{V}^{ik}\bar{U}^{mn}+
\bar{U}^{ik}\bar{V}^{mn}+\bar{M}^{ik}\bar{M}^{mn}) -{}\nonumber\\ {}-\bar{\Psi}_3(\bar{V}^{ik}\bar{M}^{mn}+\bar{M}^{ik}\bar{V}^{mn})
-\bar{\Psi}_4\bar{V}^{ik}\bar{V}^{mn}\,.
\label{Petrov4}
\end{gather}
Five complex scalars $\Psi_0$, $\Psi_1$, $\Psi_2$, $\Psi_3$, and $\Psi_4$ are of the form
\begin{gather}
\Psi_0=-q_3W_{pqrs}l^pm^ql^rm^s,
\quad \Psi_1=-q_3W_{pqrs}l^pn^ql^rm^s,
\quad
\Psi_2=-q_3W_{pqrs}l^pm^q\bar{m}^rn^s,\nonumber\\
\Psi_3=-q_3W_{pqrs}l^pn^q\bar{m}^rn^s,
\quad \Psi_4=-q_3W_{pqrs}n^p\bar{m}^qn^r\bar{m}^s,
\label{Petrov59}
\end{gather}
and the bar above the scalar symbols denotes the complex conjugation.

According to the Petrov classification the Weyl-type tensor belongs to one of six types: $\bf I$, $\bf II$, $\bf III$, $\bf D$, $\bf N$, or $\bf O$. The type $\bf I$ is said to be algebraically general, other types are known as algebraically special. In the case of the type $\bf O$, all scalars are equal to zero, $\Psi_0=\ldots =\Psi_4=0$. For the rest types there exist one or more scalars, which can be converted into zero by the appropriate admissible turn of the tetrad vectors. The simplest set of the scalars corresponding to each type takes the form (see \cite{Exact})

\noindent
a) $\Psi_0=\Psi_1=\Psi_2=\Psi_3=\Psi_4=0$ for the type $\bf O$; \\
b) $\Psi_0=\Psi_1=\Psi_2=\Psi_3=0$, $\Psi_4=-2$ for the type $\bf N$; \\
c) $\Psi_0=\Psi_1=\Psi_3=\Psi_4=0$, $\Psi_2\neq0$ for the type $\bf D$; \\
d) $\Psi_0=\Psi_1=\Psi_2=\Psi_4=0$, $\Psi_3=-i$ for the type $\bf III$;\\
e) $\Psi_0=\Psi_1=\Psi_3=0$, $\Psi_2\neq0$, $\Psi_4=-2$ for the type $\bf II$;\\
f) $\Psi_1=\Psi_3=0$, $\Psi_0=\Psi_4\neq0$, $\Psi_2\neq0$ for the type $\bf I$.

The type $\bf O$ corresponds to a conformally flat space-time, for which the Weyl tensor and therefore the susceptibility tensor ${\cal R}_{pqrs}$ vanish. The Tamm-Rubilar tensor
${\cal G}^{(pqrs)}$ and the Fresnel equation take the simplest
form
\begin{gather}
{\cal G}^{(pqrs)}=\eps^3\,g^{(pq}g^{rs)},\\
{\cal G}^{pqrs}K_pK_qK_rK_s=\eps^3 (K_pK^p)^2=0,\label{TypeO}
\end{gather}
and if $\eps\neq0$ the dispersion relation (\ref{TypeO}) for the type $\bf O$ actually does not differ from the vacuum case. Thus, this case can be indicated as trivial. When $\eps=0$ we deal with a degenerate case, for which the Tamm-Rubilar tensor is identically equal to zero, ${\cal G}^{(pqrs)}=0$.

\subsection{General properties of the Fresnel equation}

Before we will begin to investigate other types of the Weyl tensor, we would like to formulate some general propositions about the Fresnel equation (\ref{FrenEq}).

Let us consider the case, when the wave vector $K_p$ in (\ref{FrenEq}) is null. For the sake of simplicity we assume that $K_p=l_p$. Direct calculation yields
\begin{equation}
{\cal G}^{pqrs}l_pl_ql_rl_s=
-16\Re\left[|\Psi_0|^2 \left(\Psi_2+\frac{\eps}{4}\right) -\Psi_1^2\bar\Psi_0\right].
\end{equation}
Obviously, this expression is equal to zero, if $\Psi_0=0$. Thus, when the vector $l_p$ defines a principal null direction of the Weyl tensor (see \cite{Exact}), it satisfies the Fresnel equation. Therefore for this direction the phase velocity of light propagation is equal to the vacuum speed of light and the refractive index is equal to one.

Let us proceed to the case, when the principal null direction $l_p$ defines the position of a singularity, i.e.,
when the vector $K_p=l_p$ is a solution to (\ref{singEq})
\begin{equation}
F^s\equiv{\cal G}^{(pqrs)}l_pl_ql_r=0.
\end{equation}
Calculation of every tetrad components for $F^s$ provided $\Psi_0=0$ yields
\begin{equation}
  F_{(1)}=0,\quad F_{(2)}=-24|\Psi_1|^2\left(\Re\Psi_2-\frac{\eps}{2}\right), \quad F_{(3)}=-24\Psi_1\bar\Psi_1^2,\quad F_{(4)}=-24\bar\Psi_1\Psi_1^2.
\end{equation}
From the obtained expressions we can conclude that the principal null direction $l_p$ defines a singularity of the surface (\ref{singEq}), if and only if $\Psi_0=\Psi_1=0$, i.e., if the Weyl tensor is of an algebraically special type. For this case, the equation for the amplitude vector (\ref{main}) gives
\begin{equation}
a_s=c_1l_s+c_2m_s+c_3\bar{m}_s,
\end{equation}
where $c_1$, $c_2$, and $c_3$ are arbitrary constants. The first term corresponds to the pure gauge solution $a_s\sim K_s$, while other terms define two independent polarization for one wave vector $K_s=l_s$. Thus, for algebraically special types of the Weyl tensor, the principal null direction $l_p$ determines the direction, along which light propagates without refraction and birefringence.

%
%

\subsection{Type $\bf N$}

For the type ${\bf N}$, when $\Psi_0=\Psi_1=\Psi_2=\Psi_3=0$, and $\Psi_4=-2$, the Fresnel equation (\ref{FrenEq}) gives
\begin{equation}\label{N}
  T[K]=4\eps^3\left(K_{(1)}K_{(2)}-|K_{(3)}|^2\right)^2-16\eps\,K_{(1)}^4=0.
\end{equation}
In a bizarre case, when the trace of the effective non-minimal quasi-medium dielectric tensor $\eps=1+(6q_1+3q_2+q_3)R/6=0$, the Tamm-Rubilar tensor vanishes and the Fresnel equation is trivial. When $\eps\neq0$, this equation can be rearranged as follows
\begin{equation}\label{N2}
  T[K]=\left(2K_{(1)}K_{(2)}-2|K_{(3)}|^2+\frac{4}{\eps}\,K_{(1)}^2\right)
  \left(2K_{(1)}K_{(2)}-2|K_{(3)}|^2-\frac{4}{\eps}\,K_{(1)}^2\right)=0,
\end{equation}
therefore the quartic surface defined by (\ref{N}) splits into two quadrics. The surface possesses the only singularity at $K_{(1)}=K_{(3)}=K_{(4)}=0$, i.e., at $K_p=l_p$, where both sheets of the quartic have a common point.

Each multiplier in (\ref{N2}) can be rewritten in the form $\mathfrak{g}^{pq}K_pK_q$, where the tensor $\mathfrak{g}^{pq}$ is said to be an optical metric tensor. Solutions to the Fresnel equation, in this case, have to satisfy to a relation of the following type $\mathfrak{g}^{pq}K_pK_q=0$. It means that any solution $K_p$ is a null vector for an appropriate optical metric tensor. For the type $\bf N$, we have two types of the optical metrics, for instance, $A$ and $B$,
\begin{gather}
\mathfrak{g}^{pq}_{A}=g^{pq}+\frac{4}{\eps}\,l^pl^q=l^pn^q+l^qn^p-m^p\bar{m}^q-m^q\bar{m}^p+\frac{4}{\eps}\,l^pl^q, \nonumber \\
\mathfrak{g}^{pq}_{B}=g^{pq}-\frac{4}{\eps}\,l^pl^q=l^pn^q+l^qn^p-m^p\bar{m}^q-m^q\bar{m}^p-\frac{4}{\eps}\,l^pl^q,
\label{NAB}
\end{gather}
which relate to the corresponding types of different polarizations $a^A_s$ and $a_s^B$. These formulas generalize the result obtained in \cite{BDZ2008X} for gravitational pp-waves. The birefringence is absent for the wave vector $K_p$ associated with the principal null direction $l_p$. For other directions of the wave vector, one can derive the following polarization vector expressions as solutions to Eq.(\ref{main})
\begin{gather}
a^A_s=i\left[(m_p-\bar{m}_p)l_s-(m_s-\bar{m}_s)l_p\right]K^p=i\left(V_{sp}-\bar{V}_{sp}\right)K^p,\\
a^B_s=\left[(m_p+\bar{m}_p)l_s-(m_s+\bar{m}_s)l_p\right]K^p=\left(V_{sp}+\bar{V}_{sp}\right)K^p,
\end{gather}
or
\begin{gather}
a^A_{(1)}=0,\ a^A_{(2)}=i(K_{(3)}-K_{(4)}),\ a^A_{(3)}=-iK_{(1)},\ a^A_{(4)}=iK_{(1)},\\
a^B_{(1)}=0,\ a^B_{(2)}=K_{(3)}+K_{(4)},\ a^B_{(3)}=K_{(1)},\ a^B_{(4)}=K_{(1)}.
\end{gather}
Both these vectors are real, non-null and orthogonal to each other and to the wave vector:
\begin{gather}
g^{pq}a_p^Aa_q^A=g^{pq}a_p^Ba_q^B=-2(K_pl^p)^2\neq0,\quad g^{pq}a_p^Aa_q^B=0,\quad a_p^AK^p=a_p^BK^p=0.
\end{gather}

\subsection{Type {\bf III}}

For the type ${\bf III}$, when $\Psi_0=\Psi_1=\Psi_2=\Psi_4=0$, and $\Psi_3=-i$, the dispersion relation (\ref{FrenEq}) gives
\begin{equation}\label{III}
  4\eps^3\left(K_{(1)}K_{(2)}-|K_{(3)}|^2\right)^2-16\eps\,K_{(1)}^2\left(K_{(1)}K_{(2)}+3|K_{(3)}|^2\right)
  +32i K_{(1)}^3\left(K_{(3)}-K_{(4)}\right)=0.
\end{equation}
In the case $\eps=0$, this equation splits into
\begin{equation}\label{IIIb}
  K_{(1)}=0,\quad K_{(3)}=K_{(4)},
\end{equation}
and describes two intersecting linear surfaces. When $\eps\neq0$, we deal with a qualitatively different situation, because the surface defined by (\ref{III}) does not split, e.g., into quadrics, but it is an essentially quartic one.
This surface possesses two sheets, which intersect at two singular points in $\mathbb{R}P^3$: first one corresponds to the principal null direction, $K_p=l_p$, and the second one is located at
\begin{equation}
K_p=n_p+\frac{9}{4\eps^2}l_p-\frac{i}{2\eps}\left(m_p-\bar{m}_p\right).
\end{equation}
Along these directions the birefringence phenomenon is absent, and the polarization vector for the latter case being orthogonal to $K_p$ takes the form
\begin{equation}
a_p=C_1\left[\frac12 l_p-\frac{2\eps^2}{3}n_p+i\eps(m_p-\bar{m}_p)\right]+C_2(m_p+\bar{m}_p).
\end{equation}

\subsection{Type {\bf D}}

When we deal with the type ${\bf D}$, i.e., if $\Psi_0=\Psi_1=\Psi_3=\Psi_4=0$, and $\Psi_2\neq0$, like for the type $\bf N$, we have a bizarre case, namely, $\Re\Psi_2=\eps/2$, for which the Fresnel equation is trivial. If $\Re\Psi_2\neq\eps/2$ the dispersion relation (\ref{FrenEq}) reduces to the form
\begin{equation}\label{D}
  (\eps+4\Re\Psi_2)\left(K_{(1)}K_{(2)}-|K_{(3)}|^2\right)^2-\frac{36|\Psi_2|^2}{\eps-
  2\Re\Psi_2}K_{(1)}K_{(2)}|K_{(3)}|^2=0.
\end{equation}
In the case $\Re\Psi_2=-\eps/4$, the first term vanishes and Eq.(\ref{D}) gives three linear equations, $K_{(1)}=0$, $K_{(2)}=0$, and $K_{(3)}=K_{(4)}=0$, describing three intersecting surfaces. At last, if $\Re\Psi_2\neq\eps/2$ and $\Re\Psi_2\neq -\eps/4$, we have the most interesting case, for which the fourth-order equation (\ref{D}) splits
into two second-order equations
\begin{gather}
K_{(1)}K_{(2)}-M|K_{(3)}|^2=0,\label{Da}\\
K_{(1)}K_{(2)}-\frac{1}{M}|K_{(3)}|^2=0, \label{Db}
\end{gather}
where the factor $M$ is determined as follows
\begin{equation}
  M=\frac{|\eps+2\Psi_2-\bar\Psi_2|+3|\Psi_2|}{|\eps+2\Psi_2-\bar\Psi_2|-3|\Psi_2|}.\label{M}
\end{equation}
These formulas generalize the result obtained by Drummond and Hathrell \cite{NM2} for the Schwarzschild space-time.

Like in the case of the type $\bf N$, for the type $\bf D$ we can represent two optical metrics:
\begin{gather}
\mathfrak{g}^{pq}_{A}=l^{p}n^{q}+l^{q}n^{p}-M(m^{p}{\bar{m}}^{q}+m^q{\bar{m}}^p),\nonumber \\
\mathfrak{g}^{pq}_{B}=l^{p}n^{q}+l^{q}n^{p}-M^{-1}(m^{p}{\bar{m}}^{q}+m^q{\bar{m}}^p).\label{DAB}
\end{gather}

The quartic surface describing by (\ref{D}) has two singularities, at $K_p=l_p$ and $K_p=n_p$, which are common points of the quadric sheets (\ref{Da}) and (\ref{Db}) of the main surface. Here the vector $n^p$ is another principal null direction of the type $\bf D$ Weyl tensor. When $K_i\neq l_i$ and $K_i\neq
n_i$, there exist two different polarizations related to the corresponding optical metric
\begin{gather}
a^A_i=K_p\left[S(l^pn_i-n^pl_i)+m^p\bar{m}_i-\bar{m}^pm_i\right],\\
a^B_i=K_p\left[l^pn_i-n^pl_i+S(m^p\bar{m}_i-\bar{m}^pm_i)\right],
\end{gather}
where the factor $S$ takes the form
\begin{equation}
S=\frac{i}{\Im\Psi_2}\frac{|\eps+2\Psi_2-\bar\Psi_2|\Re\Psi_2-(\eps+\Re\Psi_2)|\Psi_2|}
{|\eps+2\Psi_2-\bar\Psi_2|+3|\Psi_2|}.
\end{equation}
When an imagine part of $\Psi_2$ tends to zero, $S$ is finite and vanishes, if $\Re\Psi_2\neq-\eps$.

\subsection{Type II}

For the last algebraically special type ${\bf II}$, when $\Psi_0=\Psi_1=\Psi_3=0$, $\Psi_2\neq0$, and $\Psi_4=-2$, the Fresnel equation takes the form
\begin{align}\label{II}
  T[K]&=4(\eps-2\Re\Psi_2)^2(\eps+4\Re\Psi_2)\left(K_{(1)}K_{(2)}-|K_{(3)}|^2\right)^2 \nonumber\\
  {}&-144(\eps-2\Re\Psi_2)|\Psi_2|^2K_{(1)}K_{(2)}|K_{(3)}|^2-16(\eps+4\Re\Psi_2)K_{(1)}^4 \nonumber\\
  {}&+48(\eps+2\bar{\Psi}_2-\Psi_2)\Psi_2 K_{(1)}^2K_{(4)}^2+48(\eps-\bar{\Psi}_2+2\Psi_2)\bar{\Psi}_2 K_{(1)}^2K_{(3)}^2=0.
\end{align}
Obviously, this equation reduces to (\ref{N}) for the type $\bf N$, when we put $\Psi_2=0$.

For the first specific case, $\eps=2\Re\Psi_2$, we obtain
\begin{equation}\label{II1}
    -48K_{(1)}^2\left[2\Re\Psi_2K_{(1)}^2-3|\Psi_2|^2(K_{(3)}^2+K_{(4)}^2)\right]=0.
\end{equation}
This equation describes two intersecting surfaces, those are of the first and the second order, respectively.
For the second specific case, $\eps=-4\Re\Psi_2$, the relation (\ref{II}) yields
\begin{equation}\label{II2}
    -144K_{(1)}
    \left[-3\Re\Psi_2  |\Psi_2|^2K_{(2)}|K_{(3)}|^2+K_{(1)}(\Psi_2^2K_{(4)}^2+\bar{\Psi}_2^2K_{(3)}^2)\right]=0.
\end{equation}
Here the surface $T[K]=0$ splits into two parts, one of them, $K_{(1)}=0$ is of the first order, another is of the third order.

For other cases, when $\eps\neq 2\Re\Psi_2$ and $\eps\neq -4\Re\Psi_2$, the surface $T[K]=0$ does not split into lower order surfaces. However, it possesses three singularities: the first one is located at $K_{(1)}=K_{(3)}=K_{(4)}=0$ and the second and the third singularities are defined by the relations
\begin{gather}
  K_{(1)}=(\eps-2\Re\Psi_2)\sgn(\eps+4\Re\Psi_2),\quad K_{(2)}=\frac{|\eps-\bar{\Psi}_2+2\Psi_2|^2+9|\Psi_2|^2}{3|\Psi_2||\eps-\bar{\Psi}_2+2\Psi_2|},\nonumber\\
  K_{(3)}^2=(\bar{K}_{(4)})^2=\frac{(\eps-2\Re\Psi_2)^2(\eps+4\Re\Psi_2)}{3\bar\Psi_2 (\eps-\bar{\Psi}_2+2\Psi_2)}. \label{singII}
\end{gather}
These two points differ from each other by sign of $K_{(3)}$ and therefore $K_{(4)}$.

\subsection{Type I}\label{IVg}

Now let us proceed to discussion of the algebraically general type $\bf I$. For this case, we can define the set of the Weyl scalars as follows (see \cite{Exact})
\begin{equation}
\Psi_1=\Psi_3=0, \quad \Psi_0=\Psi_4=\frac{\lambda_2-\lambda_1}{2}, \quad \Psi_2=-\frac{\lambda_3}{2},
\end{equation}
where three invariants $\lambda_1$, $\lambda_2$, and $\lambda_3$ satisfy the relations
\begin{gather}
 \lambda_1+\lambda_2+\lambda_3=0,\\
 \lambda_1\lambda_2+\lambda_2\lambda_3+\lambda_3\lambda_1=-\frac{q_3^2}{16}\left(W_{ikmn}W^{ikmn}-iW_{ikmn}\starr{W}{*}^{ikmn}\right)=-I,\\
 \lambda_1\lambda_2\lambda_3=-\frac{q_3^3}{48}\left(W_{ikmn}W^{mnpq}{W^{ik}}_{pq}
 -iW_{ikmn}W^{mnpq}{\starr{W}{*}^{ik}}_{pq}\right)=2J,
\end{gather}
i.e., these invariants are the roots of the cubic equation $\lambda^3-I\lambda-2J=0$. All these quantities have to be different, because if any pair of them coincides (or, $I^3=27J^2$) the type $\bf I$ transforms to the type $\bf D$.

The Fresnel equation can be written as follows
\begin{gather}
  \beta_3\left[K_{(1)}^4+K_{(2)}^4+K_{(3)}^4+K_{(4)}^4\right]-2(\beta_3+2\alpha)
  \left[K_{(1)}^2K_{(2)}^2+|K_{(3)}|^4\right] \nonumber \\ {}+
  2(\beta_1-\beta_2+i\gamma)\left[K_{(2)}^2K_{(3)}^2+K_{(1)}^2K_{(4)}^2\right]+
   2(\beta_1-\beta_2-i\gamma)\left[K_{(2)}^2K_{(4)}^2+K_{(1)}^2K_{(3)}^2\right] \nonumber \\ {}+8(\beta_1+\beta_2+\alpha)K_{(1)}K_{(2)}|K_{(3)}|^2=0, \label{surfacek}
\end{gather}
where the quantities $\alpha$, $\beta_1$, $\beta_2$, $\beta_3$, and $\gamma$ are related to the invariants $\lambda_1$, $\lambda_2$, $\lambda_3$,
\begin{gather}
\alpha=(\eps-2\Re\lambda_1)(\eps-2\Re\lambda_2)(\eps-2\Re\lambda_3), \label{alphaX}\\
\beta_1=(\eps-2\Re\lambda_1)|\lambda_2-\lambda_3|^2, \label{beta1X}\\
\beta_2=(\eps-2\Re\lambda_2)|\lambda_3-\lambda_1|^2, \label{beta2X}\\
\beta_3=(\eps-2\Re\lambda_3)|\lambda_1-\lambda_2|^2, \label{beta3X}
\end{gather}
\begin{align}
\gamma &=-\frac{2}{3}\Im\left\{(\lambda_1-\lambda_2)[\eps+(\lambda_2-\lambda_3)](\bar\lambda_3-\bar\lambda_1){}\right.\nonumber\\
{}& \left.{}+[\eps+(\lambda_1-\lambda_2)](\bar\lambda_2-\bar\lambda_3)(\lambda_3-\lambda_1)
+(\bar\lambda_1-\bar\lambda_2)(\lambda_2-\lambda_3)[\eps+(\lambda_3-\lambda_1)]\right\}.\label{gammaX}
\end{align}

Firstly, let us consider the case, when none of the invariants $\lambda_i$ has the real part being equal to $\eps/2$. For this case $\alpha\neq0$ and every parameter $\beta_i$, $i=1,2,3$, is non-vanishing. We can calculate number of singularities for the quartic surface defined by (\ref{surfacek}). Let us assume this surface possesses at least one singularity at the point $\mathfrak{M}_1(\starr{K}{\circ}_{(1)},\starr{K}{\circ}_{(2)},\starr{K}{\circ}_{(3)},\starr{K}{\circ}_{(4)})$. Then Eqs.(\ref{singEq}) yield
\begin{gather}
\beta_3\starr{K}{\circ}_{(1)}^3-(\beta_3+2\alpha)\starr{K}{\circ}_{(1)}\starr{K}{\circ}_{(2)}^2+
(\beta_1-\beta_2+i\gamma)\starr{K}{\circ}_{(1)}\starr{K}{\circ}_{(4)}^2+
(\beta_1-\beta_2-i\gamma)\starr{K}{\circ}_{(1)}\starr{K}{\circ}_{(3)}^2 \nonumber \\
{}+2(\beta_1+\beta_2+\alpha)\starr{K}{\circ}_{(2)}\starr{K}{\circ}_{(3)}\starr{K}{\circ}_{(4)}=0,\\
\beta_3\starr{K}{\circ}_{(2)}^3-(\beta_3+2\alpha)\starr{K}{\circ}_{(2)}\starr{K}{\circ}_{(1)}^2+
(\beta_1-\beta_2+i\gamma)\starr{K}{\circ}_{(2)}\starr{K}{\circ}_{(3)}^2+
(\beta_1-\beta_2-i\gamma)\starr{K}{\circ}_{(2)}\starr{K}{\circ}_{(4)}^2 \nonumber \\
{}+2(\beta_1+\beta_2+\alpha)\starr{K}{\circ}_{(1)}\starr{K}{\circ}_{(3)}\starr{K}{\circ}_{(4)}=0,\\
\beta_3\starr{K}{\circ}_{(3)}^3-(\beta_3+2\alpha)\starr{K}{\circ}_{(3)}\starr{K}{\circ}_{(4)}^2+
(\beta_1-\beta_2+i\gamma)\starr{K}{\circ}_{(3)}\starr{K}{\circ}_{(2)}^2+
(\beta_1-\beta_2-i\gamma)\starr{K}{\circ}_{(3)}\starr{K}{\circ}_{(1)}^2 \nonumber \\
{}+2(\beta_1+\beta_2+\alpha)\starr{K}{\circ}_{(1)}\starr{K}{\circ}_{(2)}\starr{K}{\circ}_{(4)}=0,\\
\beta_3\starr{K}{\circ}_{(4)}^3-(\beta_3+2\alpha)\starr{K}{\circ}_{(4)}\starr{K}{\circ}_{(3)}^2+
(\beta_1-\beta_2+i\gamma)\starr{K}{\circ}_{(4)}\starr{K}{\circ}_{(1)}^2+
(\beta_1-\beta_2-i\gamma)\starr{K}{\circ}_{(4)}\starr{K}{\circ}_{(2)}^2 \nonumber \\
{}+2(\beta_1+\beta_2+\alpha)\starr{K}{\circ}_{(1)}\starr{K}{\circ}_{(2)}\starr{K}{\circ}_{(3)}=0. \label{singEqI}
\end{gather}
From these equations we obtain that the parameters $\beta_1$, $\beta_2$, $\beta_3$, and $\gamma$ are determined by $\alpha$ and the coordinates of the point $\mathfrak{M}_1$,
\begin{gather}
\beta_1=\frac{\alpha(\starr{K}{\circ}_{(1)}\starr{K}{\circ}_{(2)}-\starr{K}{\circ}_{(3)}\starr{K}{\circ}_{(4)})
(\starr{K}{\circ}_{(1)}^2+\starr{K}{\circ}_{(2)}^2-\starr{K}{\circ}_{(3)}^2-\starr{K}{\circ}_{(4)}^2)}
{2\left[\starr{K}{\circ}_{(1)}\starr{K}{\circ}_{(2)}(\starr{K}{\circ}_{(3)}^2+\starr{K}{\circ}_{(4)}^2)+
\starr{K}{\circ}_{(3)}\starr{K}{\circ}_{(4)}(\starr{K}{\circ}_{(1)}^2+\starr{K}{\circ}_{(2)}^2)\right]}, \label{beta1}\\
\beta_2=\frac{\alpha(\starr{K}{\circ}_{(1)}\starr{K}{\circ}_{(2)}-\starr{K}{\circ}_{(3)}\starr{K}{\circ}_{(4)})
(\starr{K}{\circ}_{(1)}^2+\starr{K}{\circ}_{(2)}^2+\starr{K}{\circ}_{(3)}^2+\starr{K}{\circ}_{(4)}^2)}
{2\left[\starr{K}{\circ}_{(3)}\starr{K}{\circ}_{(4)}(\starr{K}{\circ}_{(1)}^2+\starr{K}{\circ}_{(2)}^2)-
\starr{K}{\circ}_{(1)}\starr{K}{\circ}_{(2)}(\starr{K}{\circ}_{(3)}^2+\starr{K}{\circ}_{(4)}^2)\right]},  \label{beta2}\\
\beta_3=\frac{4\alpha (\starr{K}{\circ}_{(1)}^2\starr{K}{\circ}_{(2)}^2-\starr{K}{\circ}_{(3)}^2\starr{K}{\circ}_{(4)}^2)}
{(\starr{K}{\circ}_{(1)}^2-\starr{K}{\circ}_{(2)}^2)^2-(\starr{K}{\circ}_{(3)}^2-\starr{K}{\circ}_{(4)}^2)^2}, \label{beta3}\\
\gamma=\frac{i\alpha(\starr{K}{\circ}_{(1)}^2-\starr{K}{\circ}_{(2)}^2)(\starr{K}{\circ}_{(3)}^2-\starr{K}{\circ}_{(4)}^2)
(\starr{K}{\circ}_{(1)}^2\starr{K}{\circ}_{(2)}^2-\starr{K}{\circ}_{(3)}^2\starr{K}{\circ}_{(4)}^2)
\left[(\starr{K}{\circ}_{(1)}^2+\starr{K}{\circ}_{(2)}^2)^2-(\starr{K}{\circ}_{(3)}^2+\starr{K}{\circ}_{(4)}^2)^2\right]}
{\left[(\starr{K}{\circ}_{(1)}^2-\starr{K}{\circ}_{(2)}^2)^2-(\starr{K}{\circ}_{(3)}^2-\starr{K}{\circ}_{(4)}^2)^2\right]
\left[\starr{K}{\circ}_{(3)}^2\starr{K}{\circ}_{(4)}^2(\starr{K}{\circ}_{(1)}^2+\starr{K}{\circ}_{(2)}^2)^2-
\starr{K}{\circ}_{(1)}^2\starr{K}{\circ}_{(2)}^2(\starr{K}{\circ}_{(3)}^2+\starr{K}{\circ}_{(4)}^2)^2\right]}. \label{gamma}
\end{gather}
Excluding from Eqs.(\ref{beta1})-(\ref{gamma}) the coordinates $\starr{K}{\circ}_{(1)}$, \ldots, $\starr{K}{\circ}_{(4)}$, we find this singular point $\mathfrak{M}_1$ exists if
\begin{equation}\label{determ}
  \alpha(\beta_1^2+\beta_2^2+\beta_3^2+\gamma^2)-
  2\alpha(\beta_1\beta_2+\beta_2\beta_3+\beta_3\beta_1)-4\beta_1\beta_2\beta_3=0,
\end{equation}
and the parameters $\alpha$, $\beta_1$, $\beta_2$, $\beta_3$, and $\gamma$ defined by (\ref{alphaX})-(\ref{gammaX}) satisfy this constraint identically.

Due to symmetry of Eqs.(\ref{singEqI}), in addition to the point $\mathfrak{M}_1$ there exist fifteen more solutions,
\begin{gather}
\mathfrak{M}_2(\starr{K}{\circ}_{(2)},\starr{K}{\circ}_{(1)},\starr{K}{\circ}_{(4)},\starr{K}{\circ}_{(3)}),\ \mathfrak{M}_3(\starr{K}{\circ}_{(1)},\starr{K}{\circ}_{(2)},-\starr{K}{\circ}_{(3)},-\starr{K}{\circ}_{(4)}),\ \mathfrak{M}_4(\starr{K}{\circ}_{(2)},\starr{K}{\circ}_{(1)},-\starr{K}{\circ}_{(4)},-\starr{K}{\circ}_{(3)}),\nonumber\\
\mathfrak{M}_5(\starr{K}{\circ}_{(1)},-\starr{K}{\circ}_{(2)},\starr{K}{\circ}_{(3)},-\starr{K}{\circ}_{(4)}),\ \mathfrak{M}_6(-\starr{K}{\circ}_{(2)},\starr{K}{\circ}_{(1)},-\starr{K}{\circ}_{(4)},\starr{K}{\circ}_{(3)}),\ \mathfrak{M}_7(\starr{K}{\circ}_{(1)},-\starr{K}{\circ}_{(2)},-\starr{K}{\circ}_{(3)},\starr{K}{\circ}_{(4)}),\nonumber \\
\mathfrak{M}_8(-\starr{K}{\circ}_{(2)},\starr{K}{\circ}_{(1)},\starr{K}{\circ}_{(4)},-\starr{K}{\circ}_{(3)}),\ \mathfrak{M}_9(\starr{K}{\circ}_{(3)},\starr{K}{\circ}_{(4)},\starr{K}{\circ}_{(1)},\starr{K}{\circ}_{(2)}),\ \mathfrak{M}_{10}(\starr{K}{\circ}_{(4)},\starr{K}{\circ}_{(3)},\starr{K}{\circ}_{(2)},\starr{K}{\circ}_{(1)}),\nonumber \\
\mathfrak{M}_{11}(\starr{K}{\circ}_{(3)},\starr{K}{\circ}_{(4)},-\starr{K}{\circ}_{(1)},-\starr{K}{\circ}_{(2)}),\ \mathfrak{M}_{12}(\starr{K}{\circ}_{(4)},\starr{K}{\circ}_{(3)},-\starr{K}{\circ}_{(2)},\starr{K}{\circ}_{(1)}),\
\mathfrak{M}_{13}(\starr{K}{\circ}_{(3)},-\starr{K}{\circ}_{(4)},\starr{K}{\circ}_{(1)},-\starr{K}{\circ}_{(2)}),\nonumber \\ \mathfrak{M}_{14}(-\starr{K}{\circ}_{(4)},\starr{K}{\circ}_{(3)},-\starr{K}{\circ}_{(2)},\starr{K}{\circ}_{(1)}),\ \mathfrak{M}_{15}(\starr{K}{\circ}_{(3)},-\starr{K}{\circ}_{(4)},-\starr{K}{\circ}_{(1)},\starr{K}{\circ}_{(2)}),\ \mathfrak{M}_{16}(-\starr{K}{\circ}_{(4)},\starr{K}{\circ}_{(3)},\starr{K}{\circ}_{(2)},-\starr{K}{\circ}_{(1)}).
\end{gather}
Since a quartic surface may possess only 16 isolated singular points, other points do not exist. It is necessary to note that only four points, $\mathfrak{M}_1$, \ldots, $\mathfrak{M}_4$, have well-defined coordinates. For them the first pair of coordinates is real, while the second pair is complex conjugated. For the rest points, $\mathfrak{M}_5$, \ldots, $\mathfrak{M}_{16}$, this rule is violated. Thus, the quartic surface (\ref{surfacek}) has four isolated singular points.

With respect to the arrangement of these points the type $\bf I$ one can divide into two subtypes. For the first subtype, the singularities are located on a plane like for a biaxial crystal, i.e., the determinant of the coordinates of $\mathfrak{M}_1$, \ldots, $\mathfrak{M}_4$ vanishes,
\begin{equation}
\Delta=\left|\begin{array}{cccc}
                 \starr{K}{\circ}_{(1)} & \starr{K}{\circ}_{(2)} & \starr{K}{\circ}_{(3)} & \starr{K}{\circ}_{(4)} \\
                 \starr{K}{\circ}_{(2)} & \starr{K}{\circ}_{(1)} & \starr{K}{\circ}_{(4)} & \starr{K}{\circ}_{(3)} \\
                 \starr{K}{\circ}_{(1)} & \starr{K}{\circ}_{(2)} & -\starr{K}{\circ}_{(3)} & -\starr{K}{\circ}_{(4)} \\
                \starr{K}{\circ}_{(2)} & \starr{K}{\circ}_{(1)} & -\starr{K}{\circ}_{(4)} & -\starr{K}{\circ}_{(3)}
               \end{array}
\right|=4(\starr{K}{\circ}_{(1)}^2-\starr{K}{\circ}_{(2)}^2)(\starr{K}{\circ}_{(3)}^2-\starr{K}{\circ}_{(4)}^2).
\end{equation}
This subtype can be indicated as the Fresnelian subtype, ${\bf I}_{\rm F}$. The second type, for which $\Delta\neq0$, should be accordingly called ``non-Fresnelian'', and it can be denoted as ${\bf I}_{\rm NF}$.

The subtype ${\bf I}_{\rm F}$ corresponds evidently to the case $\gamma=0$. It is realized, for instance, when each invariant $\lambda_i$ is real. When $\gamma\neq0$ and $\eps\neq 2q_3\Re\lambda_i$ we deal with the non-Fresnelian subtype.

Second, if, however, only one of the invariants $\lambda_i$ satisfies the condition $\eps=2q_3\Re\lambda_i$, two parameters in (\ref{surfacek}), $\alpha$ and $\beta_i$ (for instance, $\beta_3$) vanish. The Fresnel equation for this case takes the form
\begin{gather}
  (\beta_1-\beta_2+i\gamma)\left[K_{(2)}^2K_{(3)}^2+K_{(1)}^2K_{(4)}^2\right]+
   (\beta_1-\beta_2-i\gamma)\left[K_{(2)}^2K_{(4)}^2+K_{(1)}^2K_{(3)}^2\right] +4(\beta_1+\beta_2)K_{(1)}K_{(2)}|K_{(3)}|^2=0. \label{surfacek0}
\end{gather}
The surface that is defined by (\ref{surfacek0}) is essentially quartic one and has two singular lines,
$K_{(1)}=K_{(2)}=0$ and $K_{(3)}=K_{(4)}=0$.

Third, if two invariants satisfy the condition $\eps=2\Re\lambda_i$, one more parameter, e.g., $\beta_2$, is equal to zero as well and the Fresnel equation can be written as follows
\begin{gather}
  (\beta_1+i\gamma)\left[K_{(2)}K_{(3)}+K_{(1)}K_{(4)}\right]^2+
   (\beta_1-i\gamma)\left[K_{(2)}K_{(4)}+K_{(1)}K_{(3)}\right]^2=0. \label{surfacek00}
\end{gather}
This surface has four singular lines:
$K_{(1)}=K_{(2)}=0$, $K_{(3)}=K_{(4)}=0$, and in addition $\left\{\begin{array}{l} K_{(1)}=K_{(2)}, \\ K_{(3)}=-K_{(4)} \end{array}\right.$ and $\left\{\begin{array}{l} K_{(1)}=-K_{(2)}, \\ K_{(3)}=K_{(4)} \end{array}\right.$.

\section{Wave surfaces}\label{V}

The wave surfaces describing the propagation of light inside (quasi-)medium play a considerable role
in the investigation of refraction properties and they are an apt illustration of the dispersion relations.
In order to proceed to discussion of wave surfaces it is necessary to define new non-null tetrads. One of them $e^i_{(t)}$ has to be time-like normalized by the condition $g_{ij}e^i_{(t)}e^j_{(t)}=1$. The remaining vectors $e^i_{(x)}$, $e^i_{(y)}$, and $e^i_{(z)}$ have to be space-like and satisfying the conditions
\begin{equation}
g_{ij}e^i_{(x)}e^j_{(x)}=g_{ij}e^i_{(y)}e^j_{(y)}=g_{ij}e^i_{(z)}e^j_{(z)}=-1.
\end{equation}
The simplest way is to choose these tetrads in the following form
\begin{equation}
e^i_{(t)}=\frac{{\rm e}^{-\phi} l^i+ {\rm e}^{\phi} n^i}{\sqrt2},\quad e^i_{(x)}=\frac{{\rm e}^{\phi}n^i-{\rm e}^{-\phi}l^i}{\sqrt2},\quad e^i_{(y)}=\frac{{\rm e}^{i\psi}m^i+{\rm e}^{-i\psi}\bar{m}^i}{\sqrt2},
\quad e^i_{(z)}=\frac{{\rm e}^{i\psi}m^i-{\rm e}^{-i\psi}\bar{m}^i}{i\sqrt2},
\end{equation}
where $\phi$ and $\psi$ are arbitrary real constants ($\phi$ is related to a speed of a frame), and their values can be fixed later.
Relations between the corresponding tetrad components of the wave vector $K_i$ can be accordingly written as
\begin{gather}
\Omega\equiv K_{(t)}=\frac{{\rm e}^{-\phi}K_{(1)}+{\rm e}^{\phi}K_{(2)}}{\sqrt2},\quad K_{(x)}=\frac{{\rm e}^{\phi}K_{(2)}-{\rm e}^{-\phi}K_{(1)}}{\sqrt2},\nonumber \\ K_{(y)}=\frac{{\rm e}^{i\psi} K_{(3)}+ {\rm e}^{-i\psi} K_{(4)}}{\sqrt2},
\quad K_{(z)}=\frac{{\rm e}^{i\psi}K_{(3)}-{\rm e}^{-i\psi}K_{(4)}}{i\sqrt2},\\
K_{(1)}={\rm e}^{\phi}\frac{\Omega-K_{(x)}}{\sqrt2},\quad  K_{(2)}={\rm e}^{-\phi}\frac{\Omega+K_{(x)}}{\sqrt2},\nonumber\\
K_{(3)}={\rm e}^{-i\psi}\frac{K_{(y)}+iK_{(z)}}{\sqrt2},\quad K_{(4)}={\rm e}^{i\psi}\frac{K_{(y)}-iK_{(z)}}{\sqrt2},
\end{gather}
where the quantity $\Omega$ may be identified as a frequency.

A wave surface determines a space distribution of the refraction indices, hence we will apply the following coordinates
\begin{equation}
n_x=\frac{K_{(x)}}{\Omega},\quad n_y=\frac{K_{(y)}}{\Omega},\quad n_z=\frac{K_{(z)}}{\Omega}.
\end{equation}
If $K_i=l_i$, then $K_{(x)}=\Omega$, $K_{(y)}=K_{(z)}=0$ or $n_x=1$, $n_y=n_z=0$. In this case light propagates along the positive direction of the axis $Ox$ and its phase velocity is equal to the speed of light. When $K_i=n_i$, we have $n_x=-1$, $n_y=n_z=0$, and light propagates along the negative direction of this axis with the same velocity.

Generally speaking, such a surface is quartic one. However, for the case of the type $\bf O$ the wave surface (of course, when $\eps\neq0$) is the unit sphere
\begin{equation}
  n_x^2+n_y^2+n_z^2=1.
\end{equation}
The rest types of wave surfaces we will consider below.

\subsection{Type $\bf N$}

\begin{figure}[t]
\halign{\hfil#\hfil&\hfil#\hfil&\hfil#\hfil\cr
\parbox[b]{0.33\textwidth}{\centerline{\includegraphics[height=5cm]{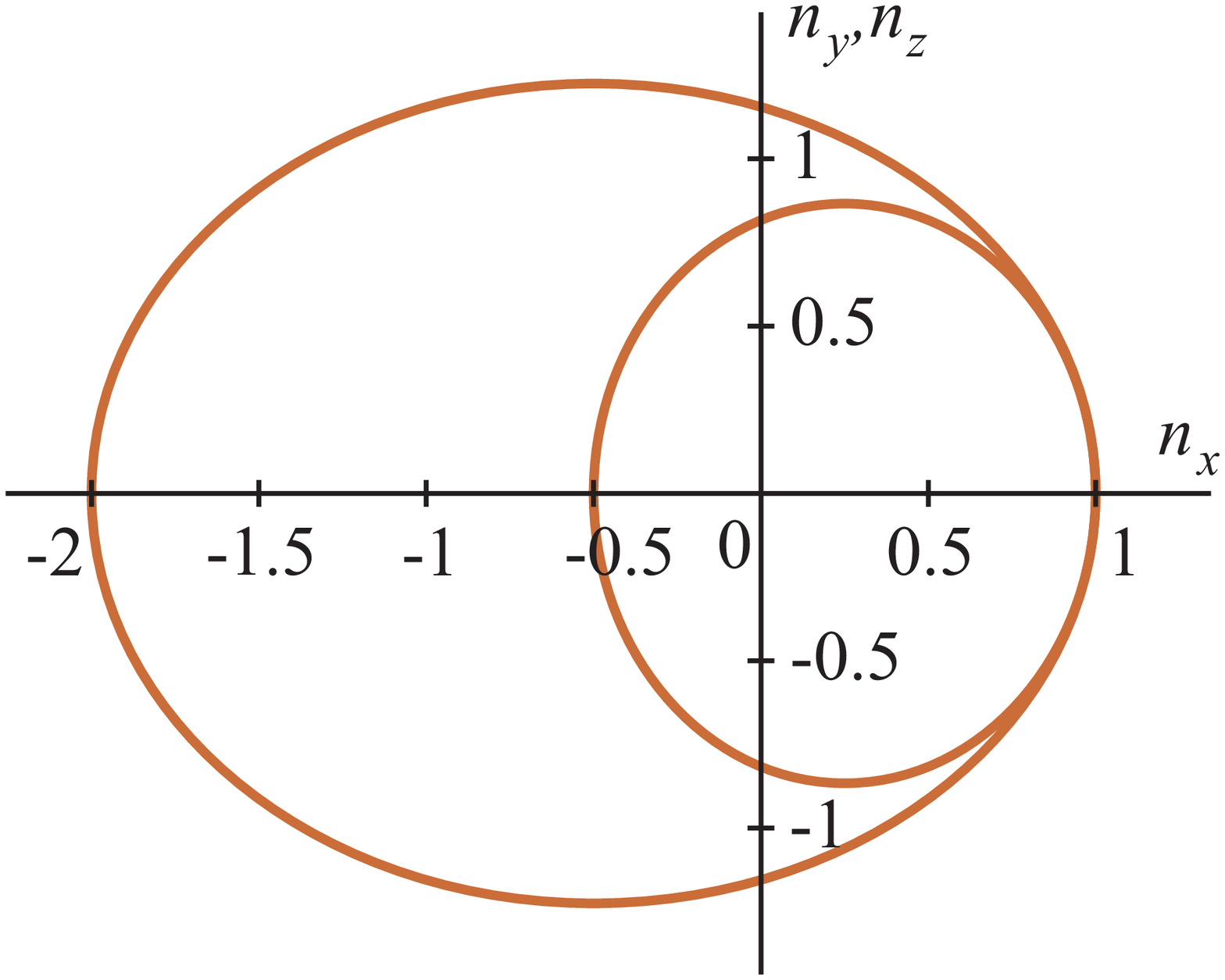}}} & \parbox[b]{0.33\textwidth}{\centerline{\includegraphics[height=5cm]{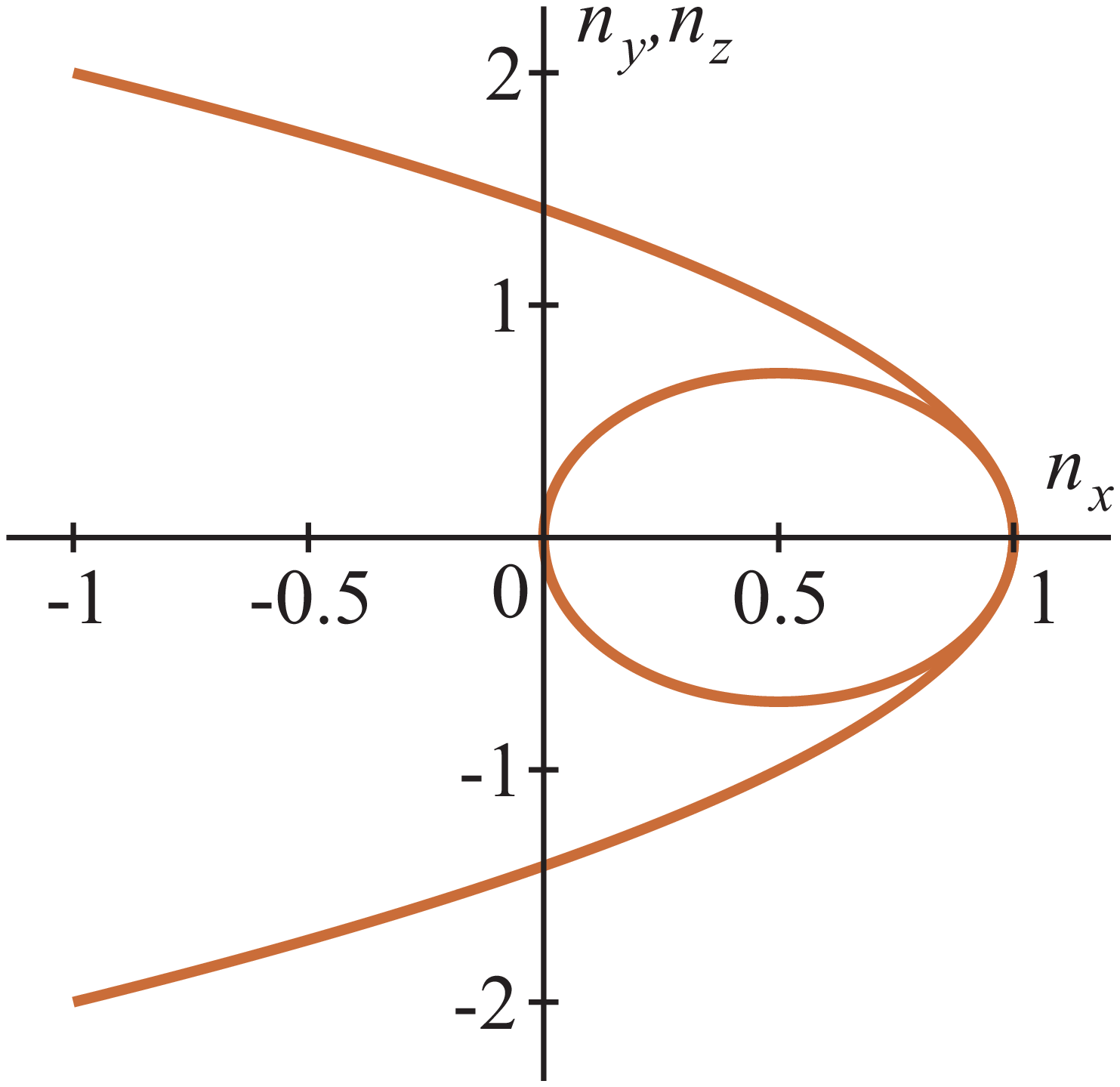}}} & \parbox[b]{0.33\textwidth}{\centerline{\includegraphics[height=5cm]{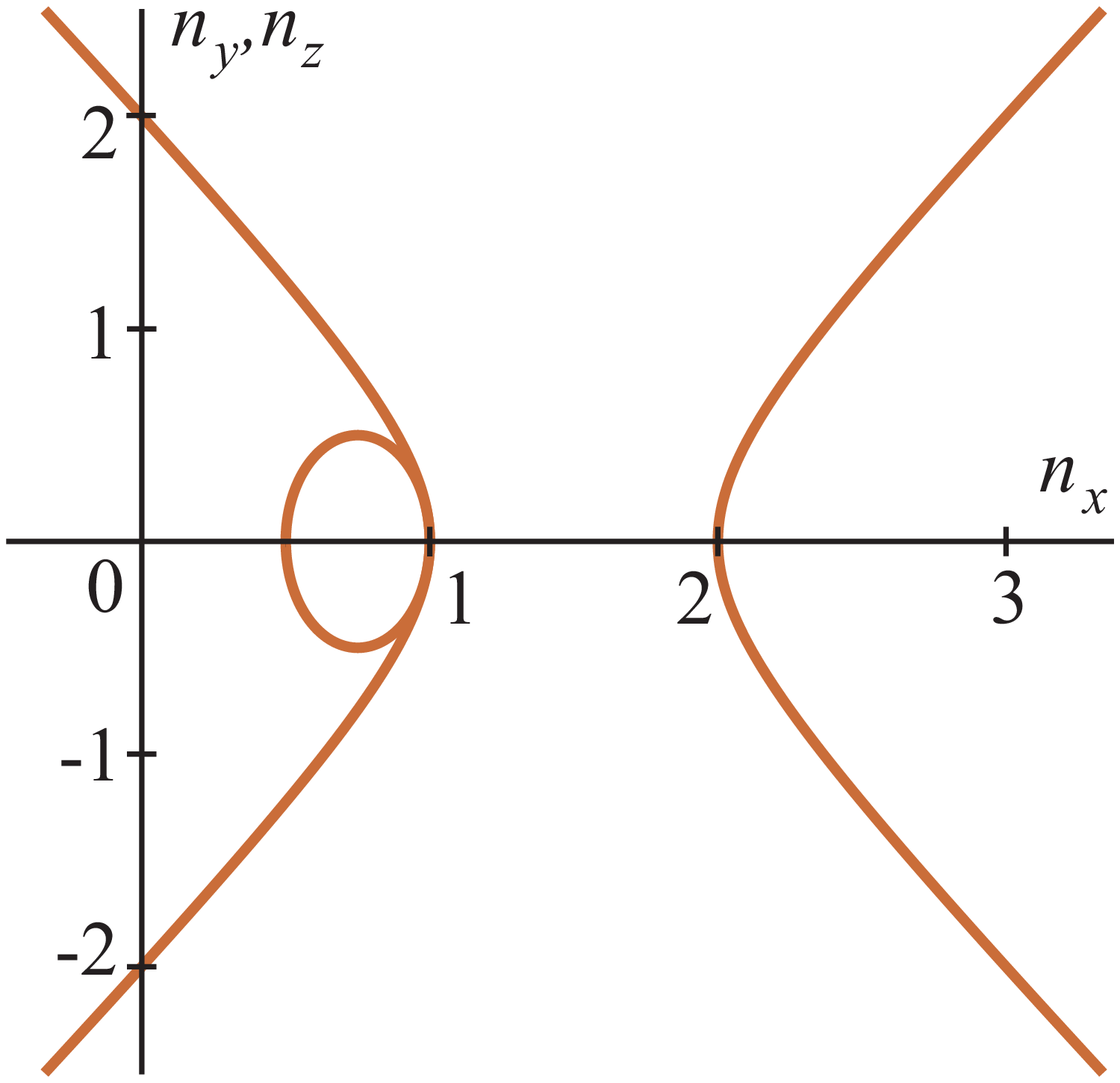}}} \cr
$(a)$ & $(b)$ & $(c)$\cr}
\caption{Axial cross-section of the wave surfaces for the type $\bf N$. The panels $(a)$, $(b)$, and $(c)$ describe
the cases $\exp(2\phi)<|\eps/2q_3|$, $\exp(2\phi)=|\eps/2|$, and $\exp(2\phi)>|\eps/2|$, respectively, the plot on the panel $(a)$ has the similar form as in \cite{Dahl2013}. All these surfaces possess the rotation symmetry.}\label{Nfig}
\end{figure}

For this type, the Fresnel equation (\ref{N2}) in the non-trivial case $\eps\neq0$ can be rewritten as follows
\begin{equation}\label{N2ws}
  \left(1-(n_x^2+n_y^2+n_z^2)+\frac{2{\rm e}^{2\phi}}{\eps}\,(1-n_x)^2\right)
  \left(1-(n_x^2+n_y^2+n_z^2)-\frac{2{\rm e}^{2\phi}}{\eps}\,(1-n_x)^2\right)=0.
\end{equation}
If we fix the value of the fitting parameter by the condition $\exp(2\phi)<|\eps/2|$, we obtain the wave surface for the type $\bf N$ splits into two ellipsoids of revolution
\begin{gather}
  \frac{\left(n_x-1+(1-2{\rm e}^{2\phi}/\eps)^{-1}\right)^2}{(1-2{\rm e}^{2\phi}/\eps)^{-2}}+\frac{n_y^2+n_z^2}{(1-2{\rm e}^{2\phi}/\eps)^{-1}}=1,\nonumber \\ \frac{\left(n_x-1+(1+2{\rm e}^{2\phi}/\eps)^{-1}\right)^2}{(1+2{\rm e}^{2\phi}/\eps)^{-2}}+\frac{n_y^2+n_z^2}{(1+2{\rm e}^{2\phi}/\eps)^{-1}}=1,
\end{gather}
which touch each other at the point $n_x=1$, $n_y=n_z=0$ (see Fig.~\ref{Nfig}a).
If we put $\exp(2\phi)=|\eps/2|$, the wave surface consists of an ellipsoid and a paraboloid of revolution (see Fig.~\ref{Nfig}b). When $\exp(2\phi)>|\eps/2|$, it splits into an ellipsoid and a two-sheet hyperboloid (see Fig.~\ref{Nfig}c).

\subsection{Type $\bf D$}

For the type $D$, as was mentioned above, there exist two specific cases, $\eps-2\Re\Psi_2=0$ and $\eps+4\Re\Psi_2=0$. The first one is trivial, while for the second case we obtain that the wave surface consists of two planes $n_x=\pm 1$ and the straight line $n_y=n_z=0$.

\begin{figure}[t]
\halign{\hfil#\hfil&\hfil#\hfil\cr
\parbox[b]{0.5\textwidth}{\centerline{\includegraphics[height=5cm]{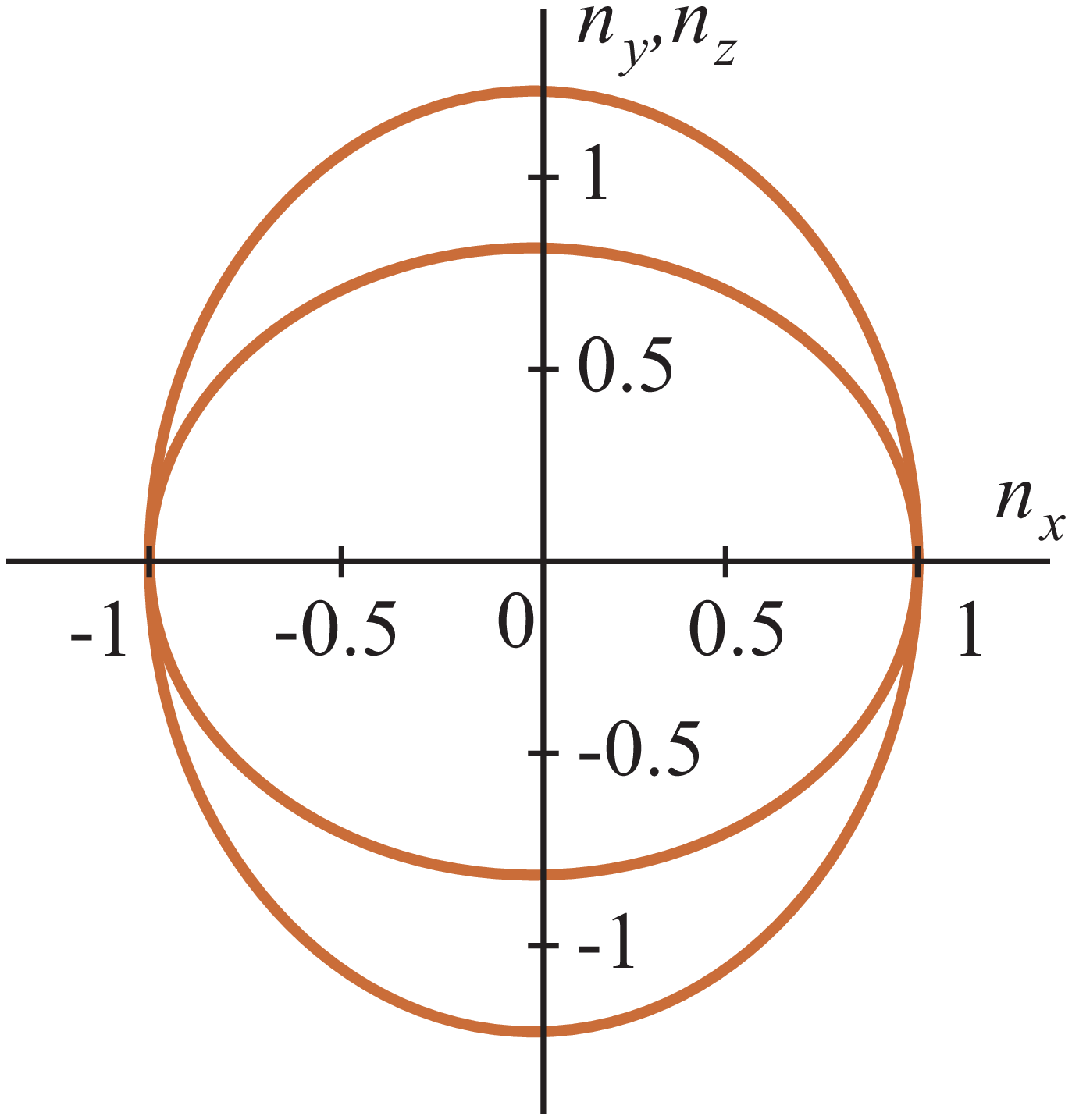}}} & \parbox[b]{0.5\textwidth}{\centerline{\includegraphics[height=5cm]{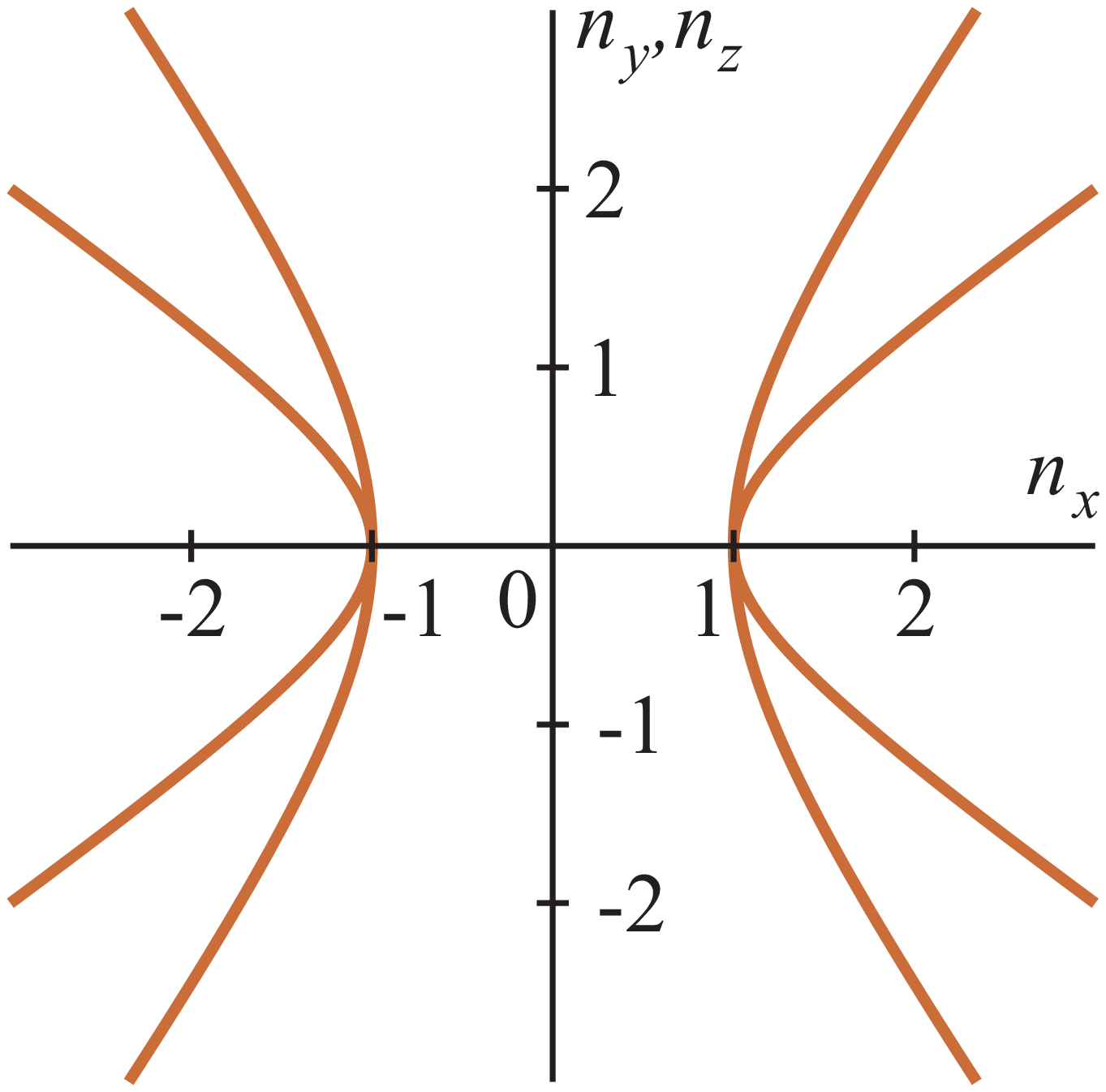}}} \cr
$(a)$ & $(b)$ \cr}
\caption{Axial cross-section of the wave surfaces for the type $\bf D$. The panels $(a)$ and $(b)$ describe
the cases $M>0$ and $M<0$, respectively. All these surfaces possess the rotation symmetry. The type $\bf D$ corresponds to the case of uniaxial quasi-medium.}\label{Dfig}
\end{figure}

When $\eps\neq 2\Re\Psi_2$ and $\eps\neq -4\Re\Psi_2$, the wave surface splits into two parts,
\begin{equation}
n_x^2+M(n_y^2+n_z^2)=1,\quad n_x^2+\frac{n_y^2+n_z^2}{M}=1, \quad
\end{equation}
which touch each other at two points, $n_x=\pm 1$, $n_y=n_z=0$. If $(\eps-2\Re\Psi_2)(\eps+4\Re\Psi_2)>0$ the quantity $M$ defined by (\ref{M}) is positive, and both parts of the wave surface are the ellipsoids of revolution (see Fig.~\ref{Dfig}a). If $(\eps-2\Re\Psi_2)(\eps+4\Re\Psi_2)<0$, we deal with two hyperboloids (see Fig.~\ref{Dfig}b). It is worth noting that in the case $\Psi_2=-\eps$ we obtain $M=-1$ and these hyperboloids coincide.

\subsection{Type $\bf III$}

\begin{figure}[h]
\halign{\hfil#\hfil&\hfil#\hfil\cr
\parbox[b]{0.5\textwidth}{\centerline{\includegraphics[height=6.5cm]{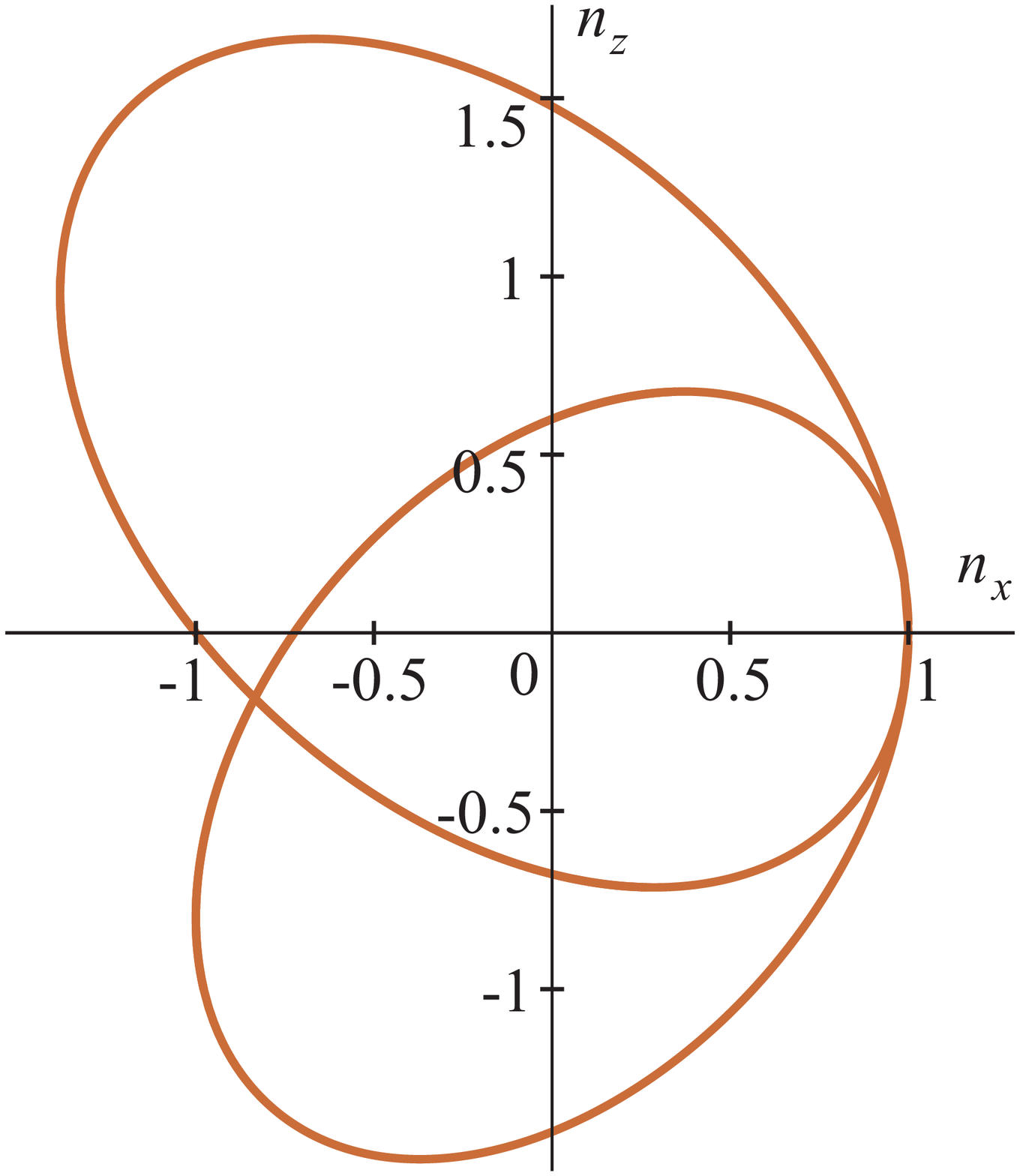}}} & \parbox[b]{0.5\textwidth}{\centerline{\includegraphics[height=6.5cm]{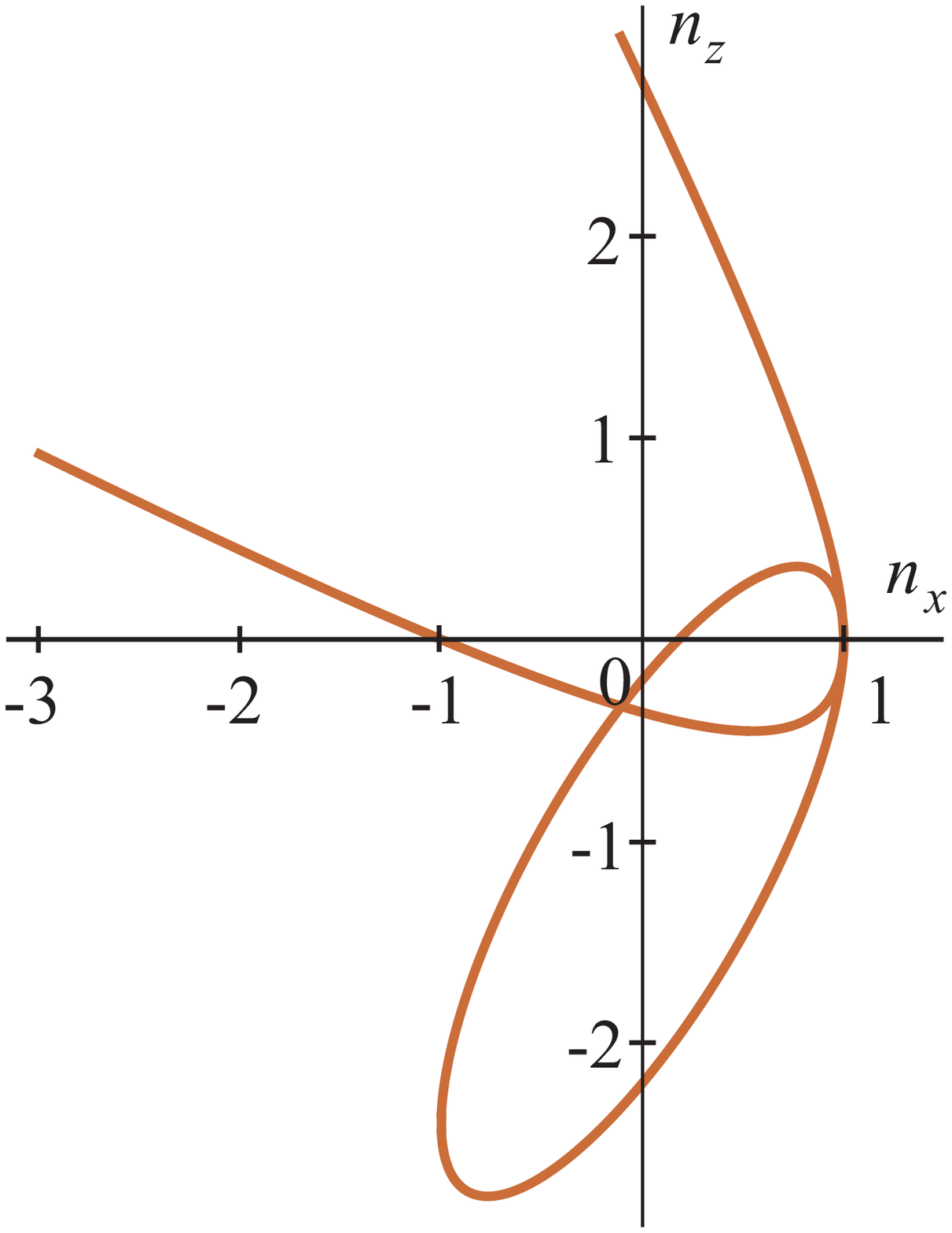}}} \cr
$(a)$ & $(b)$ \cr}
\caption{Cross-section $n_y=0$ of the wave surfaces for the type $\bf III$. On the panel $(a)$ an example for the case ${\rm e}^\phi/|\eps|<1/2$ is depicted, while the panel $(b)$ corresponds to the case ${\rm e}^\phi/|\eps|>1/2$. All these surfaces possess the mirror symmetry with respect to the plane $n_y=0$.}\label{IIIfig}
\end{figure}

For this type, the Fresnel equation (\ref{III}) in the non-trivial case $\eps\neq0$ can be rewritten as follows
\begin{gather}
(n_x^2+n_y^2+n_z^2-1)^2-\frac{4{\rm e}^{2\phi}}{\eps^2}(1-n_x)^2\left[1-n_x^2+3(n_y^2+n_z^2)\right]+\frac{16{\rm e}^{3\phi}}{\eps^3}(1-n_x)^3\left[n_y \sin\psi-n_z\cos\psi\right]=0. \label{IIIs}
\end{gather}
The surface defined by (\ref{IIIs}) is essentially quartic one and unlike the previous cases it does not split into two quadrics, e.g., ellipsoids. This wave surface (see Fig.~\ref{IIIfig}) possesses two singular points. First one, at $n_x=1$, $n_y=n_z=0$, is standard for the algebraically special types. Second one is located at the point
\begin{equation}
n_x=\frac{9{\rm e}^{2\phi}-4\eps^2}{9{\rm e}^{2\phi}+4\eps^2},\quad n_y=\frac{4{\rm e}^\phi\eps \sin\psi}{9{\rm e}^{2\phi}+4\eps^2},\quad n_z=-\frac{4{\rm e}^\phi\eps \cos\psi}{9{\rm e}^{2\phi}+4\eps^2}.
\end{equation}
Finally, the surface has the mirror symmetry with respect to the plane determined by the origin and two singular points. It appears to be finite for the case ${\rm e}^\phi/|\eps|<1/2$ (see Fig.~\ref{IIIfig}a) or infinite for the remaining values (see, e.g., Fig.~\ref{IIIfig}b).

When $\eps=0$ the wave surface transforms to the set of two planes, $n_x=1$ and $n_y\sin\psi=n_z\cos\psi$.

\subsection{Type $\bf II$}

\begin{figure}[h]
\halign{\hfil#\hfil&\hfil#\hfil\cr
\parbox[b]{0.5\textwidth}{\centerline{\includegraphics[height=6.5cm]{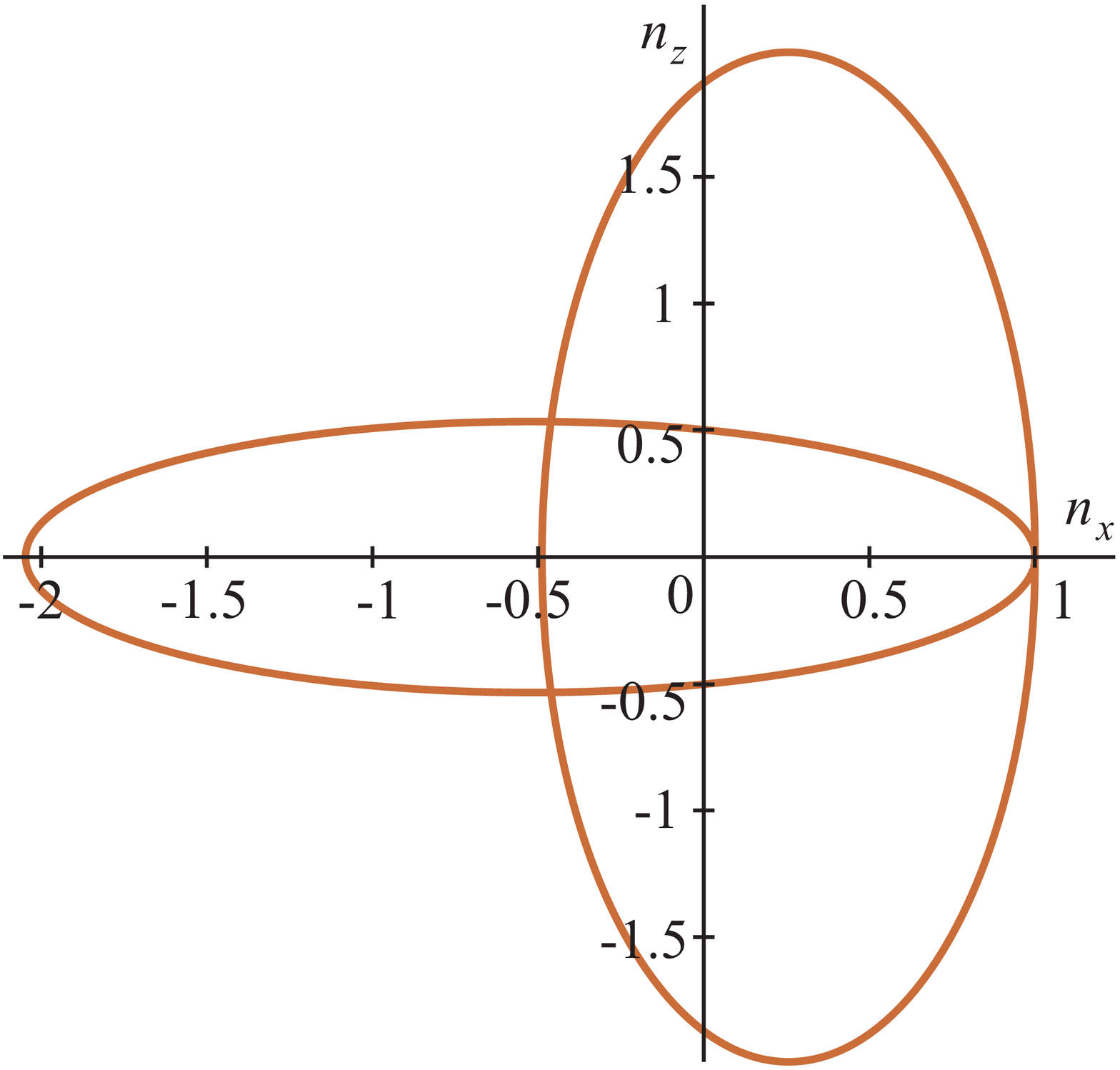}}} & \parbox[b]{0.5\textwidth}{\centerline{\includegraphics[height=6.5cm]{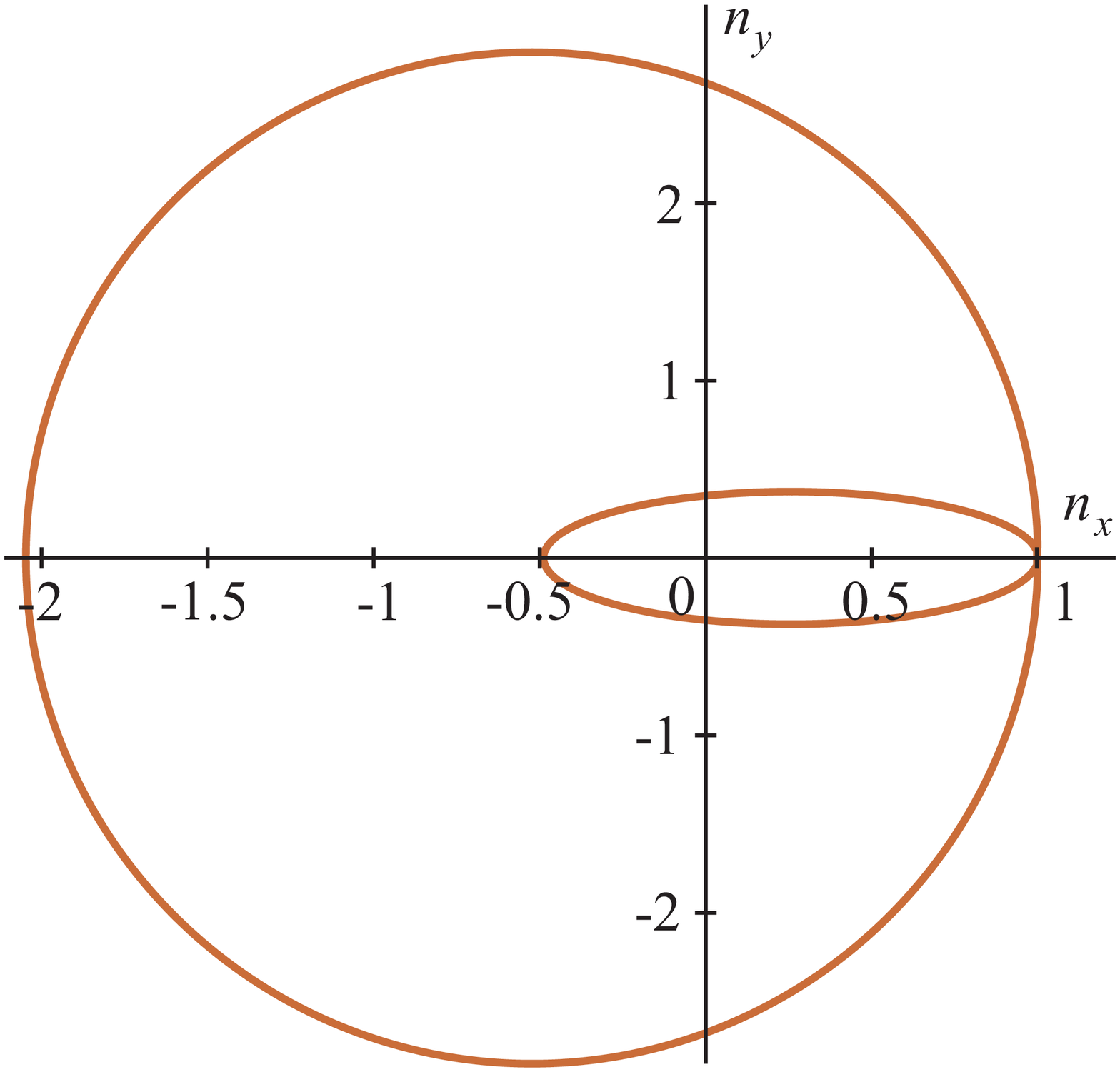}}} \cr
$(a)$ & $(b)\mathstrut $ \cr}
\caption{An example of a finite wave surface for the type $\bf II$ ($\eps=6$, $\Psi_2=0.1+1.9i$). The panels $(a)$ and $(b)$ correspond to two orthogonal cross-sections of this surface, $n_y=0$ and $n_z=0$. This surface has the mirror symmetry with respect to these planes.}\label{IIfig}
\end{figure}

\begin{figure}[h]
\halign{\hfil#\hfil&\hfil#\hfil\cr
\parbox[b]{0.5\textwidth}{\centerline{\includegraphics[height=5cm]{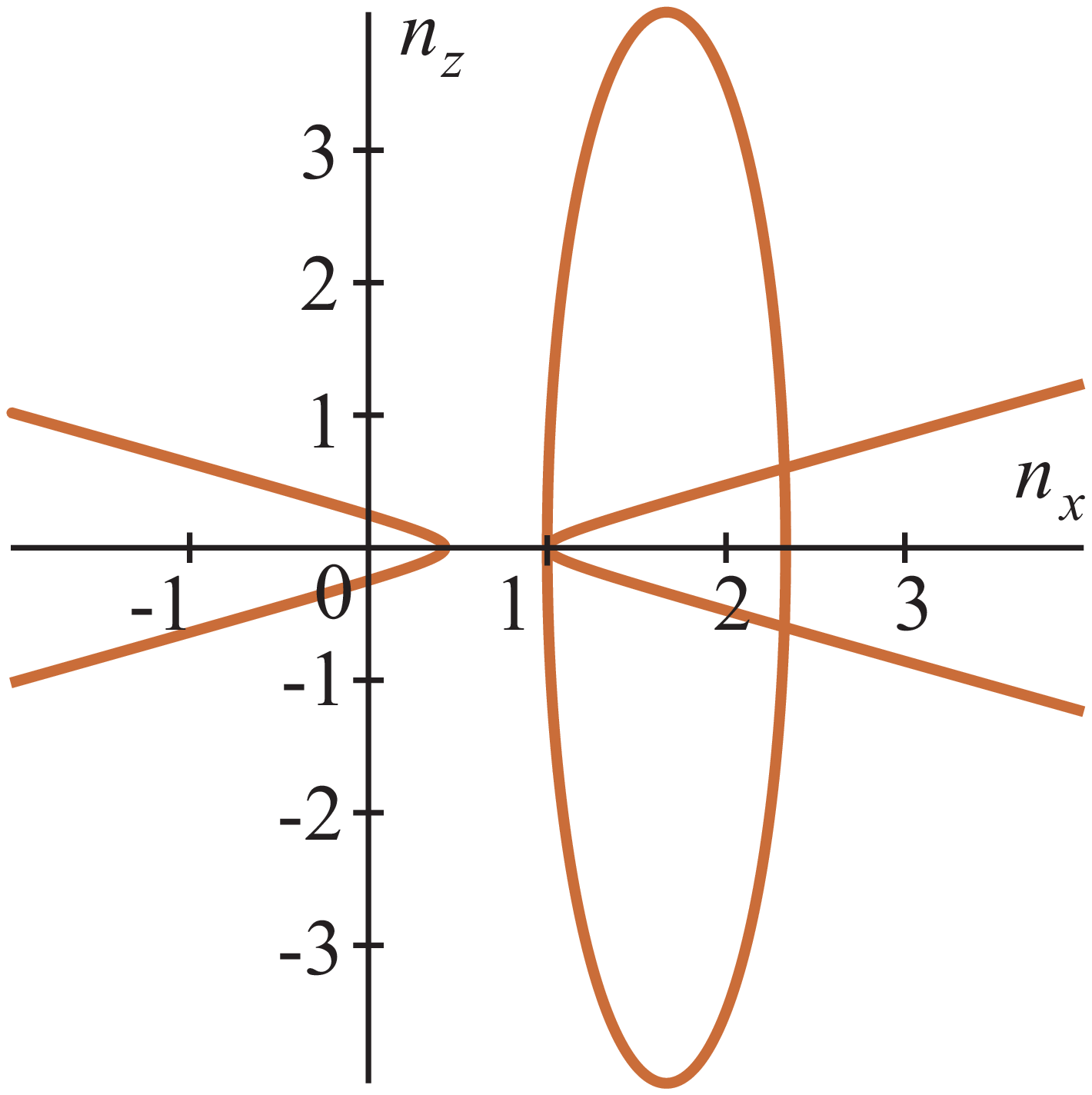}}} &
\parbox[b]{0.5\textwidth}{\centerline{\includegraphics[height=5cm]{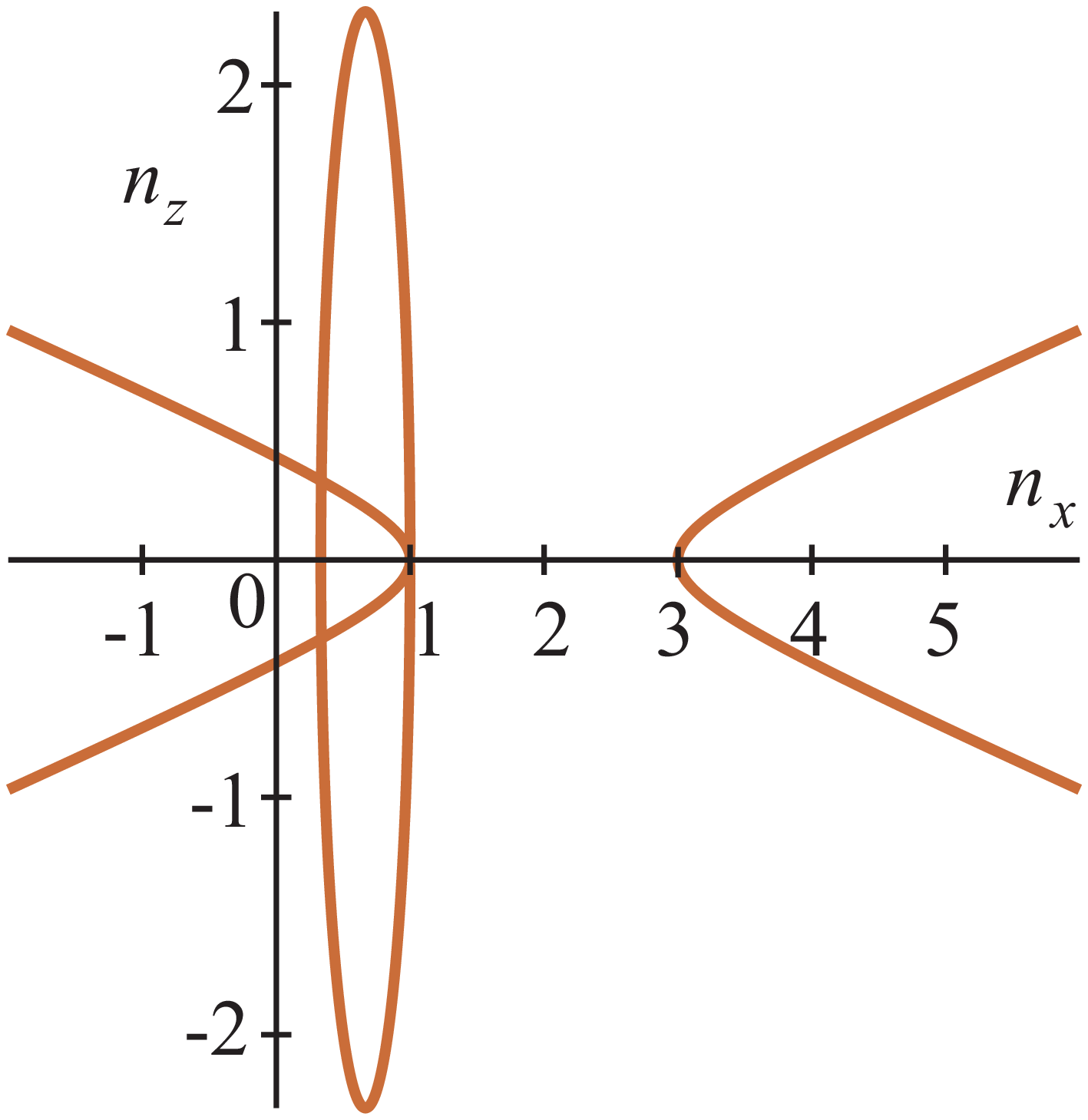}}} \cr
$(a)$ & $(b)\mathstrut $ \cr}
\vspace{5ex}
\halign{\hfil#\hfil&\hfil#\hfil\cr
\parbox[b]{0.5\textwidth}{\centerline{\includegraphics[height=5cm]{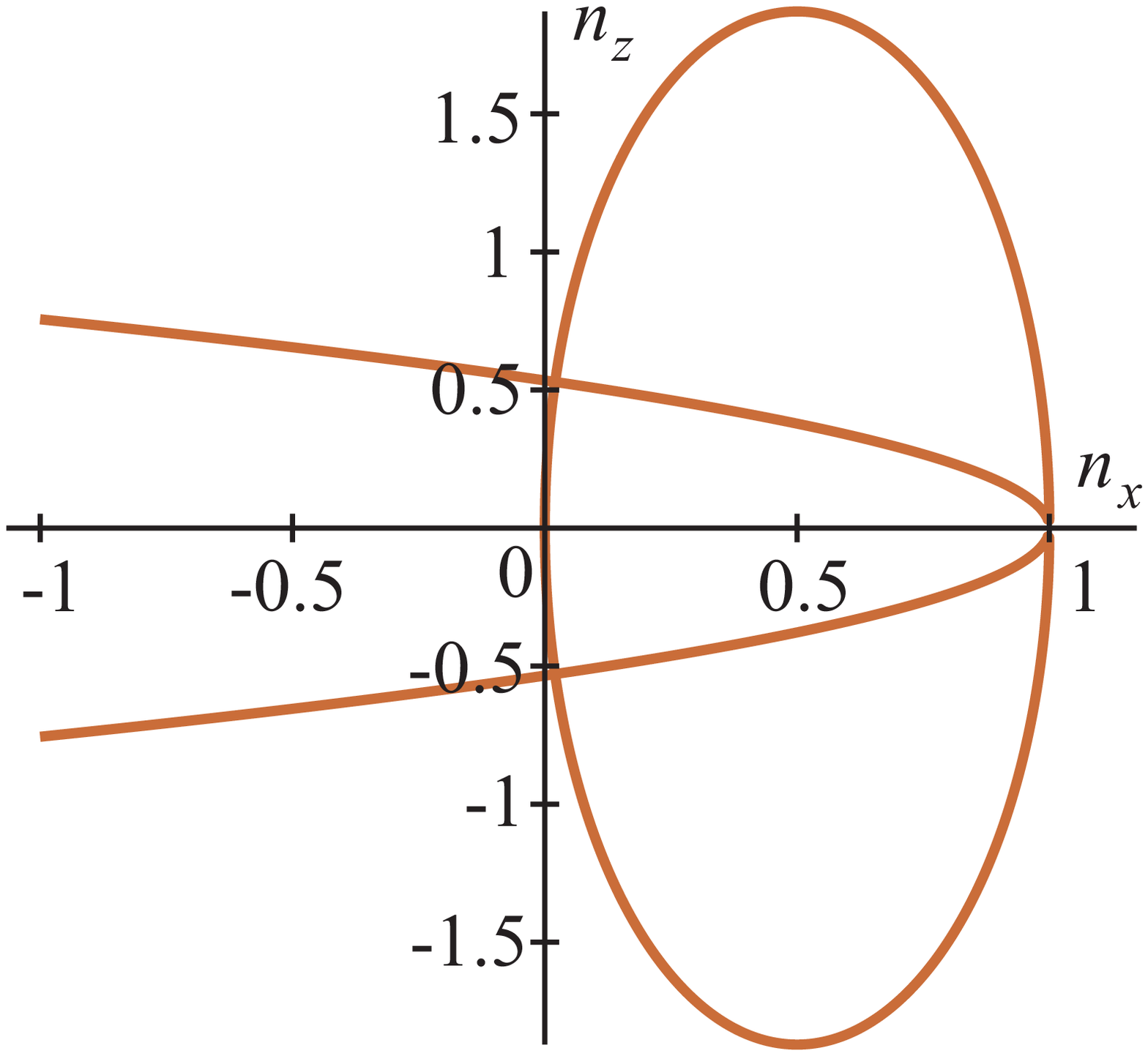}}} & \parbox[b]{0.5\textwidth}{\centerline{\includegraphics[height=5cm]{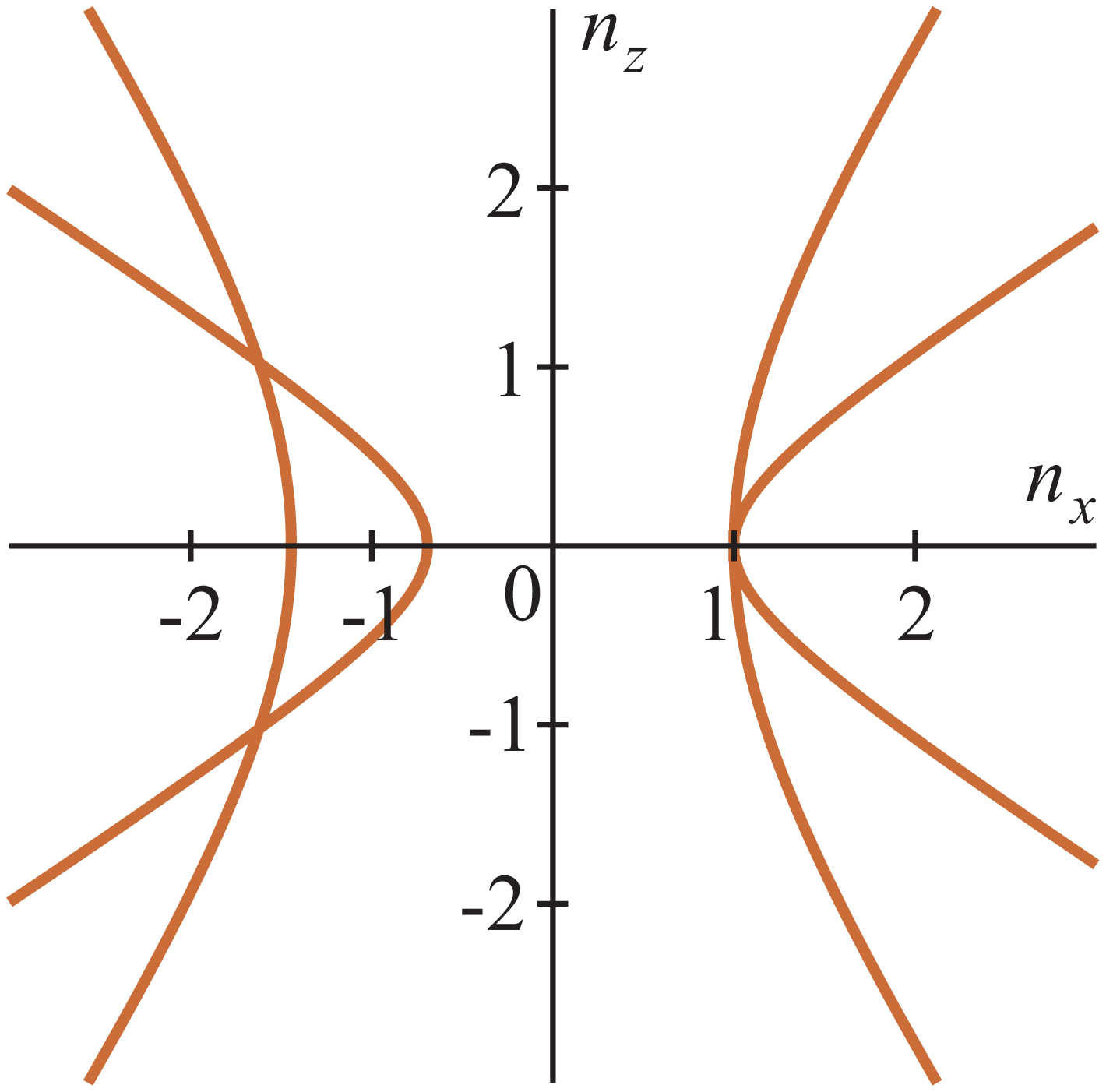}}} \cr
$(c)$ & $(d)\mathstrut $ \cr}
\caption{Examples of infinite wave surfaces for the type $\bf II$. The panels present the cross-section $n_y=0$ of such surfaces for $\eps=6$ and various values of $\Psi_2$ ($(a)$ $\Psi_2=3.4$, $(b)$ $\Psi_2=2.5$, $(c)$ $\Psi_2=2$, and $(d)$ $\Psi_2=-2.5$).}\label{IIfigX}
\end{figure}

For the type $\bf II$, the wave surface equation in terms of refraction indices takes the form
\begin{gather}
(\eps-2\Re\Psi_2)^2(\eps+4\Re\Psi_2)\left[n_x^2+n_y^2+n_z^2-1\right]^2
-36(\eps-2\Re\Psi_2)|\Psi_2|^2(1-n_x^2)(n_y^2+n_z^2) \nonumber\\
-4{\rm e}^{4\phi}(\eps+4\Re\Psi_2)(1-n_x)^4
+24{\rm e}^{2\phi}(1-n_x)^2\Re\left\{{\rm e}^{-2i\psi}\bar{\Psi}_2(\eps-\bar{\Psi}_2+2\Psi_2)(n_y+in_z)^2\right\}=0. \label{IIs}
\end{gather}
If we consider the non-trivial case, $\Re\Psi_2\neq \eps/2$ and $\Re\Psi_2\neq -\eps/4$, for the sake of simplicity, we will assume that the parameter $\psi$ is defined by the relation
\begin{equation}
\bar{\Psi}_2(\eps-\bar{\Psi}_2+2\Psi_2)=-|\Psi_2||\eps-\bar{\Psi}_2+2\Psi_2|\sgn(\eps+4\Re\Psi_2){\rm e}^{2i\psi},\label{IIpsi}
\end{equation}
which guarantees that every singularity of this surface lies on the plane $n_y=0$ (see (\ref{singII})).
For this case, the wave surface equation (\ref{IIs}) reduces to
\begin{gather}
(\eps-2\Re\Psi_2)^2(\eps+4\Re\Psi_2)\left[n_x^2+n_y^2+n_z^2-1\right]^2
-36(\eps-2\Re\Psi_2)|\Psi_2|^2(1-n_x^2)(n_y^2+n_z^2) \nonumber\\
-4{\rm e}^{4\phi}(\eps+4\Re\Psi_2)(1-n_x)^4 -24{\rm e}^{2\phi}|\Psi_2||\eps-\bar{\Psi}_2+2\Psi_2|\sgn(\eps+4\Re\Psi_2)(1-n_x)^2(n_y^2-n_z^2)=0.\label{IIsX}
\end{gather}
This equation describes an essentially quartic surface as well as for the type $\bf III$. It possesses the mirror symmetry with respect to the planes $n_y=0$ and $n_z=0$ and three singular points: one of them is located at $n_x=1$, $n_y=n_z=0$, position of two remaining points is determined by (\ref{singII}) with (\ref{IIpsi}). This surface appears to be finite (see Fig.~\ref{IIfig}) only when
\begin{equation}
\left\{\begin{array}{l}
(\eps-2\Re\Psi_2)^2>4{\rm e}^{4\phi},\\
{\rm e}^{2\phi}>\dfrac{(\eps-2\Re\Psi_2)^2|\eps+4\Re\Psi_2|}{12|\Psi_2||\eps-\bar{\Psi}_2+2\Psi_2|}\cdot\dfrac{|\eps-\bar{\Psi}_2+2\Psi_2|^2+9|\Psi_2|^2}{9|\Psi|^2-|\eps-\bar{\Psi}_2+2\Psi_2|^2}.
\end{array}\right.,
\end{equation}
otherwise, it will be infinite (see possible examples in Figs.~\ref{IIfigX}).


For the first specific case, $\eps=2\Re\Psi_2\neq0$, the equation (\ref{IIsX}) can be transformed to the form
\begin{equation}
(1-n_x)^2\left[{\rm e}^{2\phi}|\Re\Psi_2|(1-n_x)^2 +3|\Psi_2|^2(n_y^2-n_z^2)\right]=0.
\end{equation}
It describes a surface consisted of two parts: a plane $n_x=1$ and a cone ${\rm e}^{2\phi}|\Re\Psi_2|(1-n_x)^2 +3|\Psi_2|^2(n_y^2-n_z^2)=0$, whose vertex is located at the point $n_x=1$, $n_y=n_z=0$.

For the second specific case, $\eps=-4\Re\Psi_2\neq0$, the wave surface equation (\ref{IIs}) can be rewritten as follows
\begin{gather}
3\Re\Psi_2|\Psi_2|^2(1-n_x^2)(n_y^2+n_z^2)
-{\rm e}^{2\phi}(1-n_x)^2\Re\left\{{\rm e}^{-2i\psi}\bar{\Psi}_2^2(n_y+in_z)^2\right\}=0. \label{IIsXX}
\end{gather}
As it was done for the non-trivial case, we can choose the value of the parameter $\psi$ to simplify this expression. We assume that $\bar\Psi_2=|\Psi_2|{\rm e}^{i\psi}$. It allows to reduce (\ref{IIsXX}) to the form
\begin{gather}
(1-n_x)\left[3\Re\Psi_2(1+n_x)(n_y^2+n_z^2)
-{\rm e}^{2\phi}(1-n_x)(n_y^2-n_z^2)\right]=0. \label{IIsXXX}
\end{gather}
The surface defined by this equation splits into two parts: a plane $n_x=1$ and a cubic surface
\begin{equation}
3\Re\Psi_2{\rm e}^{-2\phi}(1+n_x)(n_y^2+n_z^2)
-(1-n_x)(n_y^2-n_z^2)=0.
\end{equation}

When $\eps=0$ and $\Re\Psi_2=0$ the wave surface for both previous cases degenerates to a set of planes: $n_x=1$, $n_y=\pm n_z$.

\subsection{Type $\bf I$}

\begin{figure}[h]
\halign{\hfil#\hfil&\hfil#\hfil\cr
\parbox[b]{0.5\textwidth}{\centerline{\includegraphics[height=5cm]{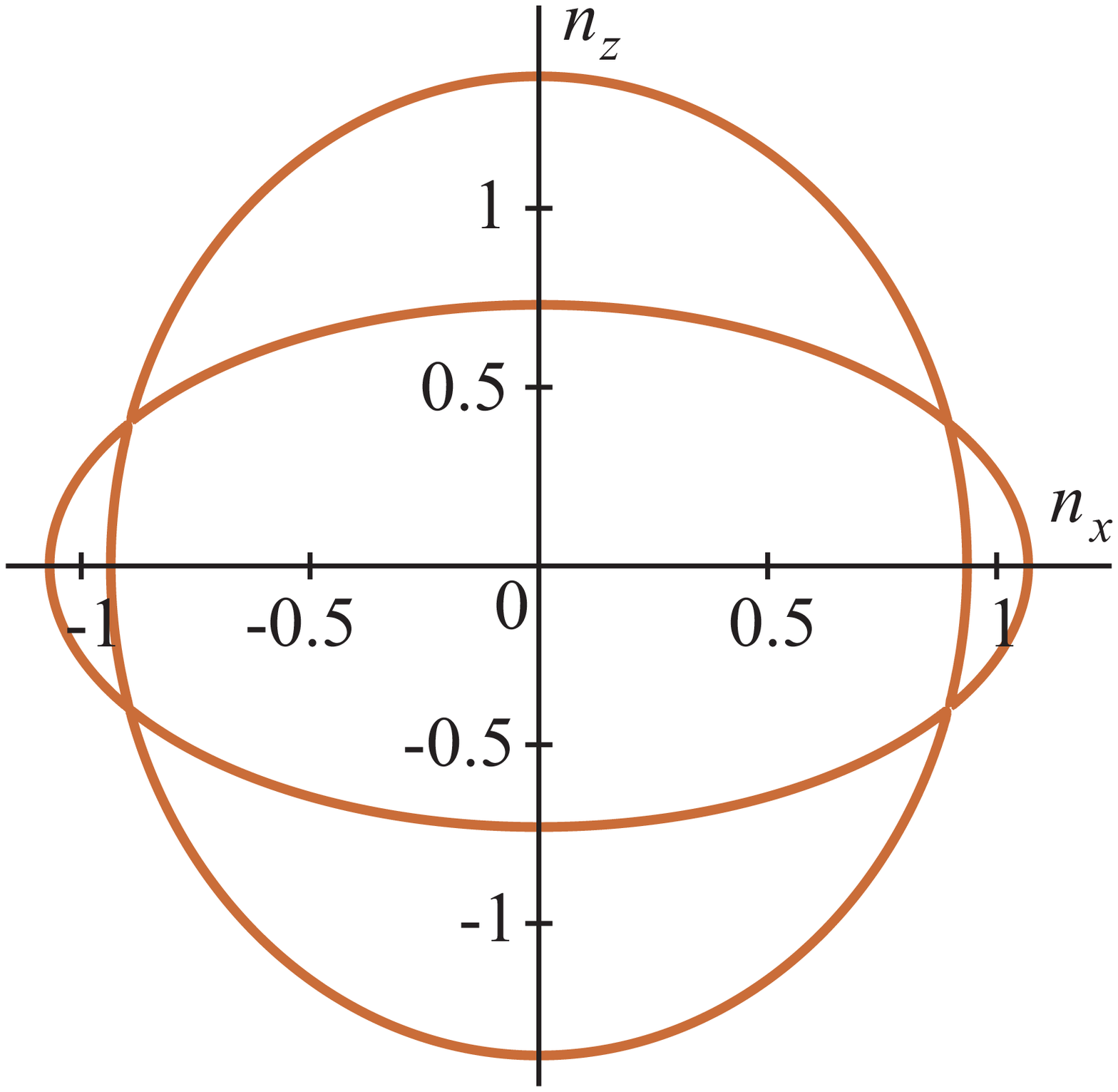}}} & \parbox[b]{0.5\textwidth}{\centerline{\includegraphics[height=5cm]{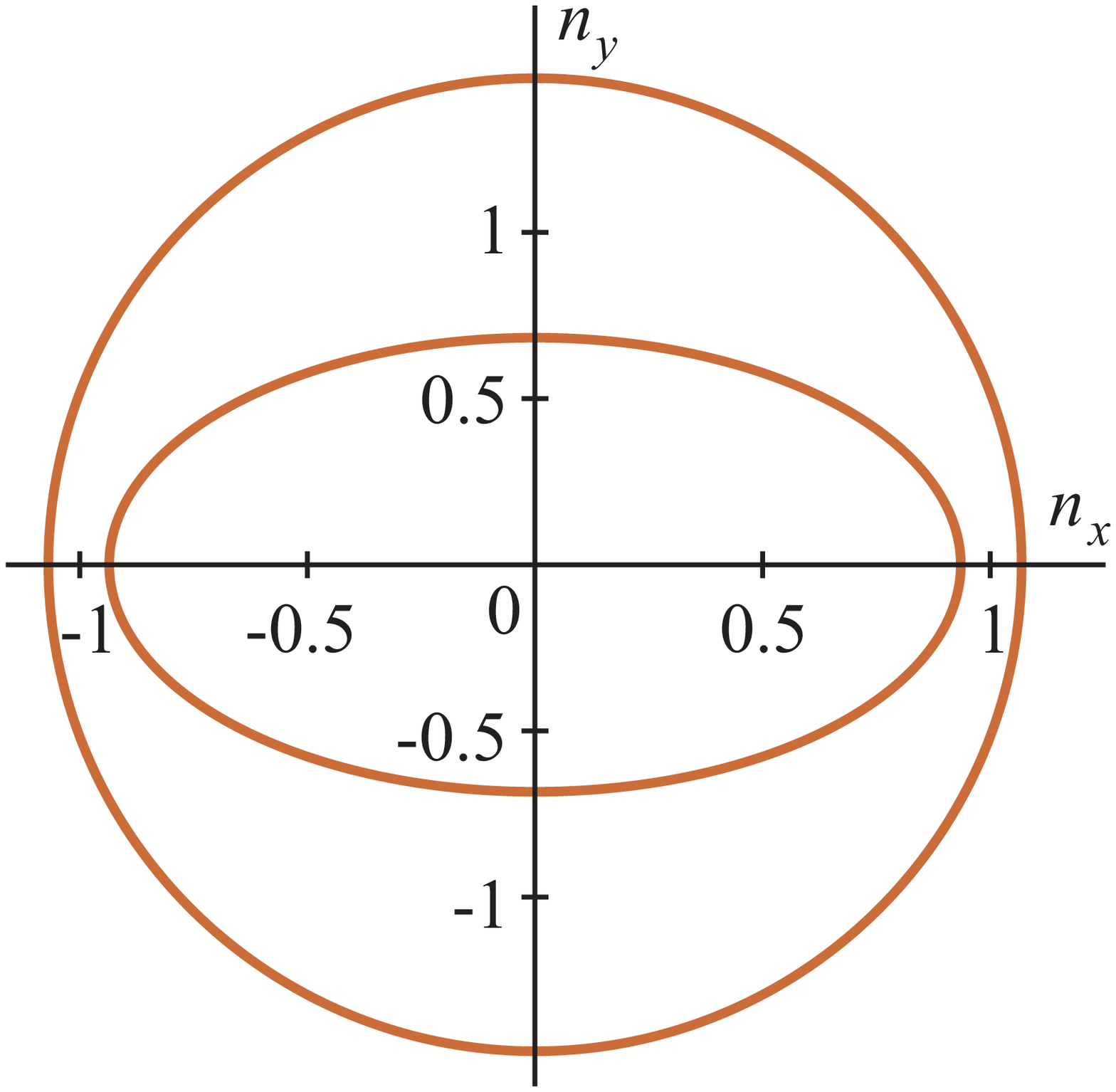}}} \cr
$(a)$ & $(b)\mathstrut $ \cr}
\caption{An example of a finite wave surface for the Fresnelian subtype ${\bf I}_{\rm F}$ ($\eps=1$, $\lambda_1=0.1$, $\lambda_2=0.15$). The panels $(a)$ and $(b)$ correspond to two orthogonal cross-sections of this surface, $n_y=0$ and $n_z=0$. This surface has the mirror symmetry with respect to these planes and all four singular points lies on one plane $n_y=0$. This subtype corresponds to the case of biaxial quasi-medium.}\label{IFfig}
\end{figure}

\begin{figure}[h]
\halign{\hfil#\hfil&\hfil#\hfil&\hfil#\hfil\cr
\parbox[b]{0.33\textwidth}{\centerline{\includegraphics[height=5cm]{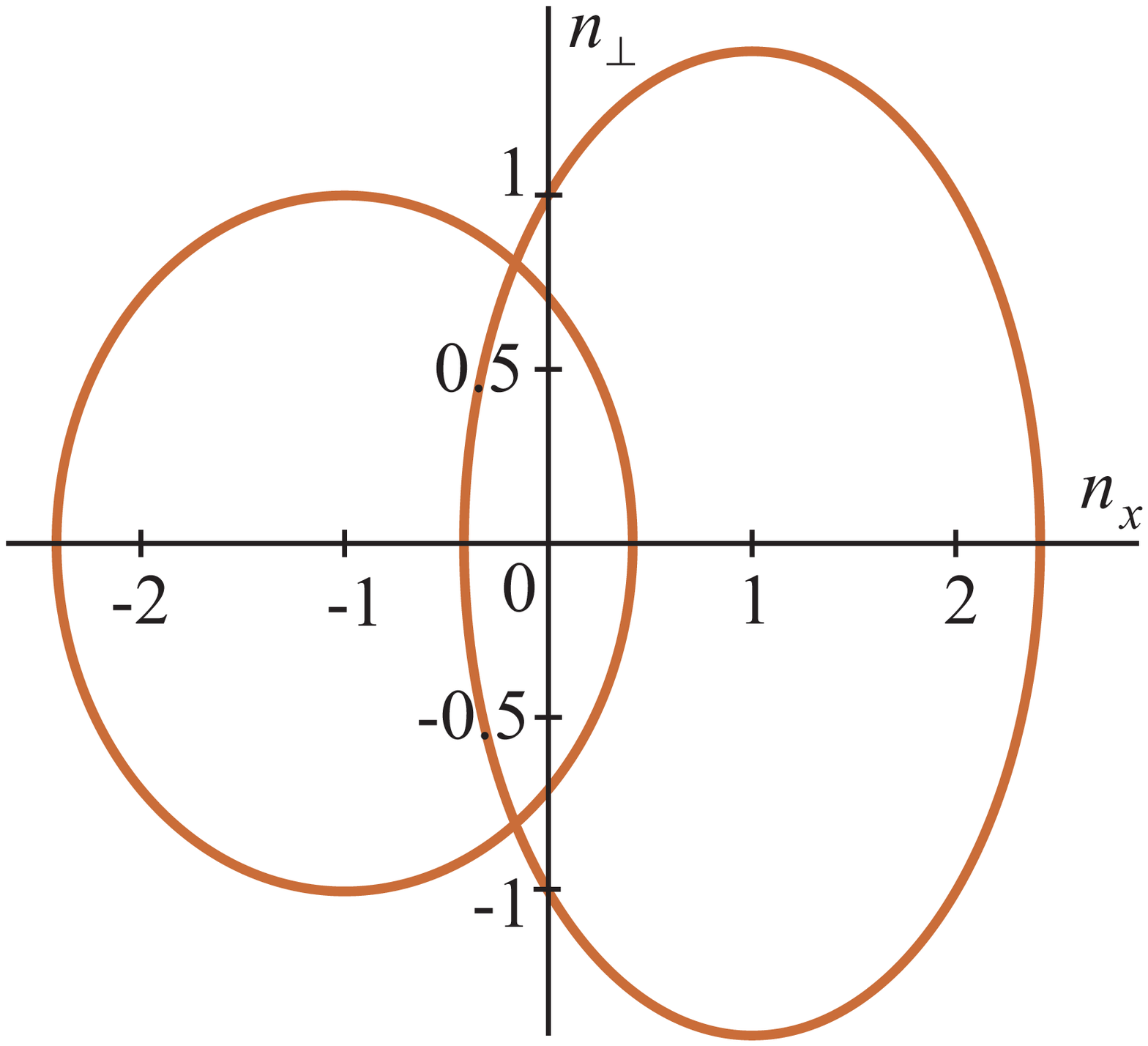}}} & \parbox[b]{0.33\textwidth}{\centerline{\includegraphics[height=5cm]{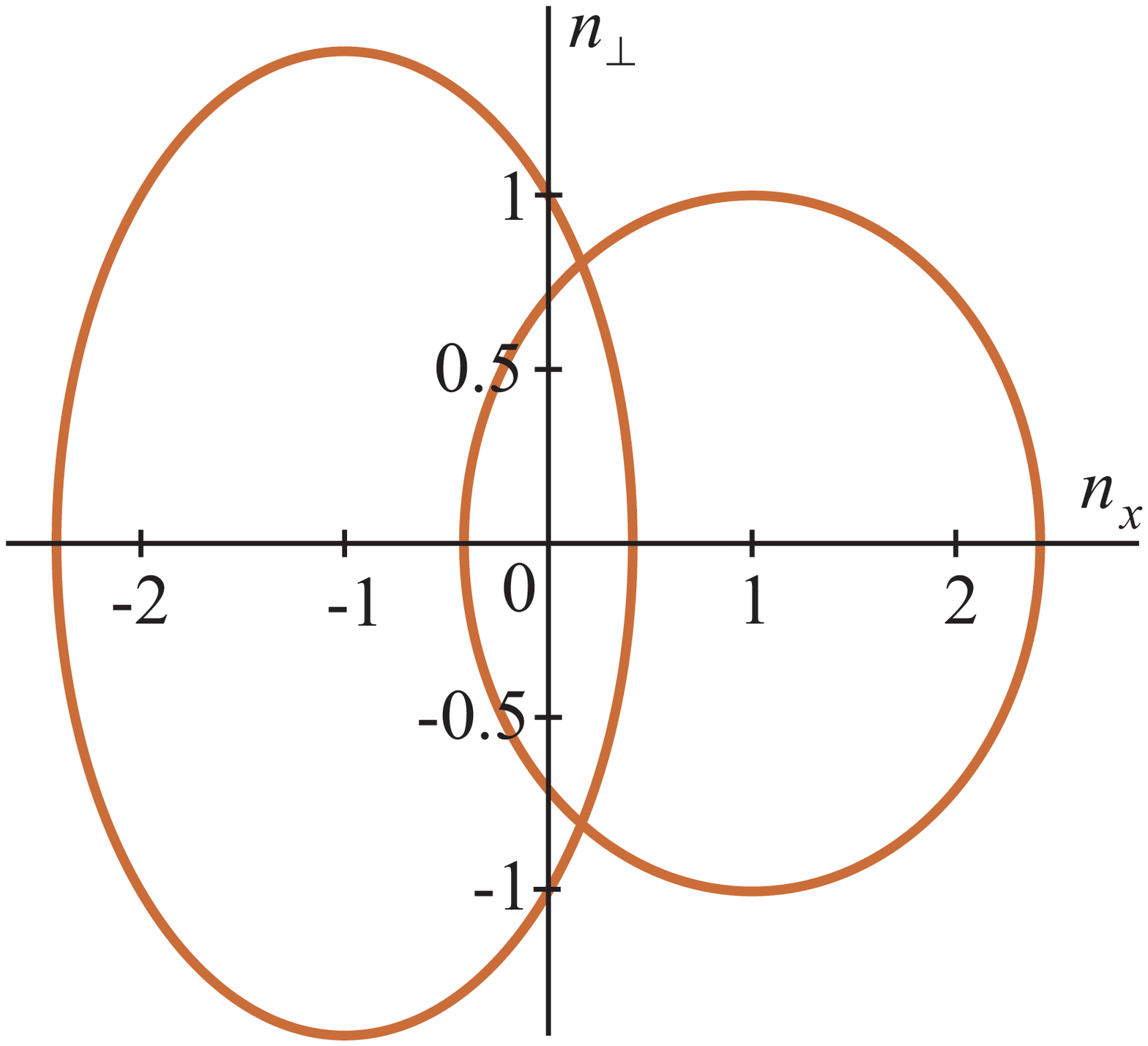}}} &
\parbox[b]{0.33\textwidth}{\centerline{\includegraphics[height=5cm]{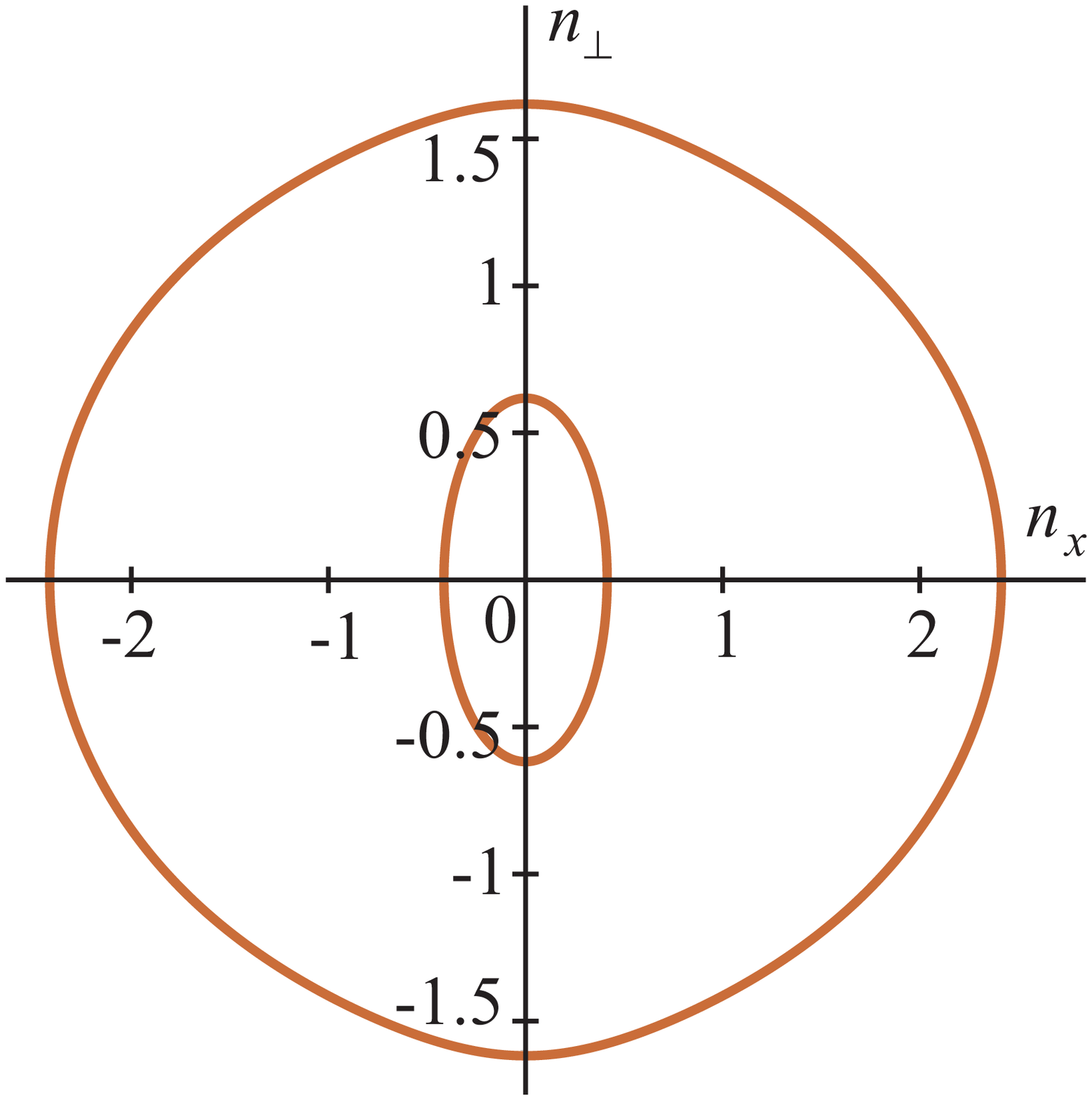}}} \cr
$(a)$ & $(b)\mathstrut $ & $(c)$ \cr}
\caption{An example of a finite wave surface for the non-Fresnelian subtype ${\bf I}_{\rm NF}$ ($\eps=1$, $\lambda_1=0.5i$, $\lambda_2=-0.5i$). The panels $(a)$ and $(b)$ correspond to cross-sections with inclination to the plane $n_z=0$ being equal to $\pi/4$ and $-\pi/4$, respectively. The panel $(c)$ describes the cross-section $n_y=0$.}\label{INFfig}
\end{figure}

\begin{figure}[h]
\halign{\hfil#\hfil&\hfil#\hfil&\hfil#\hfil\cr
\parbox[b]{0.33\textwidth}{\centerline{\includegraphics[height=5cm]{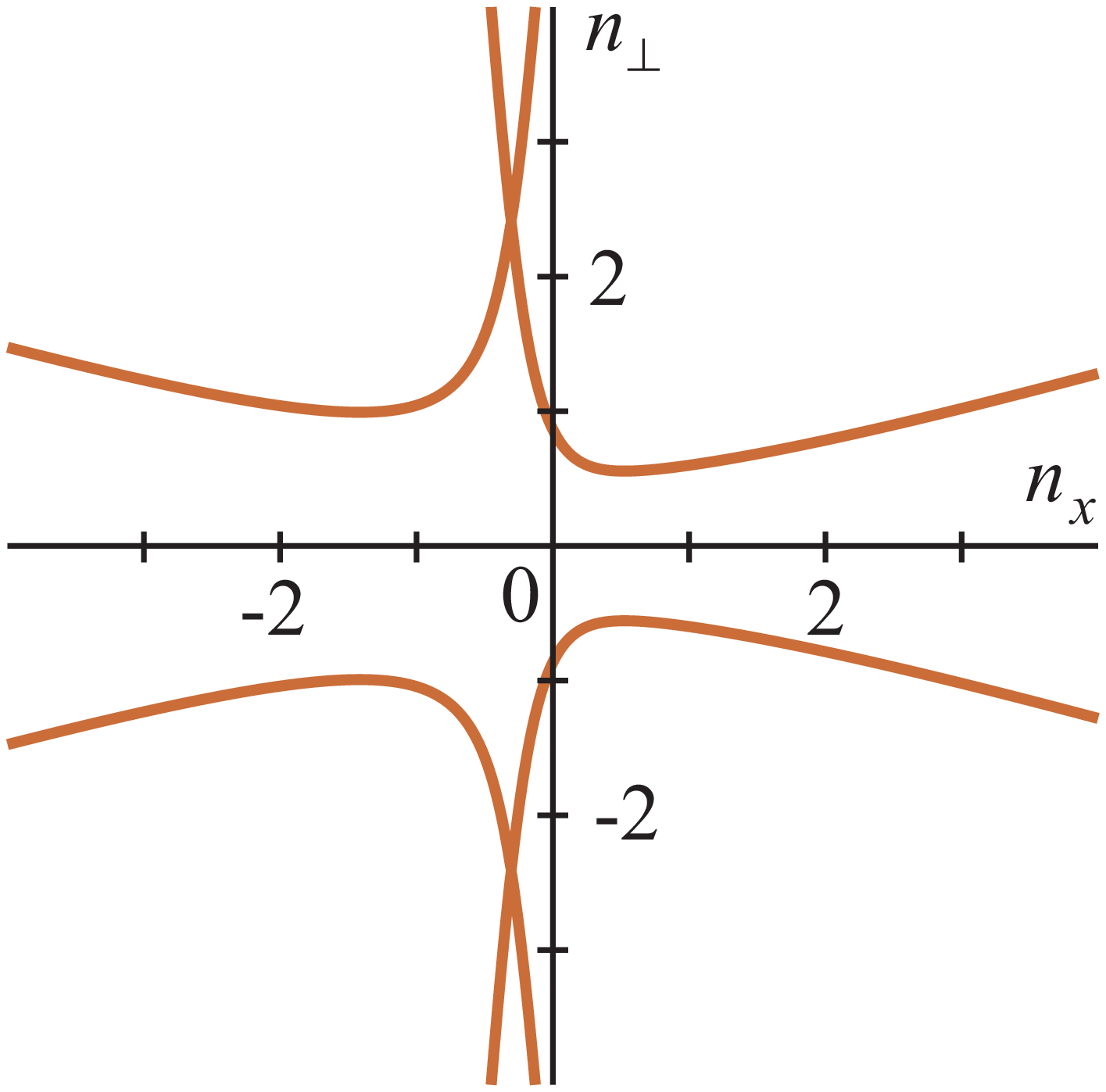}}} & \parbox[b]{0.33\textwidth}{\centerline{\includegraphics[height=5cm]{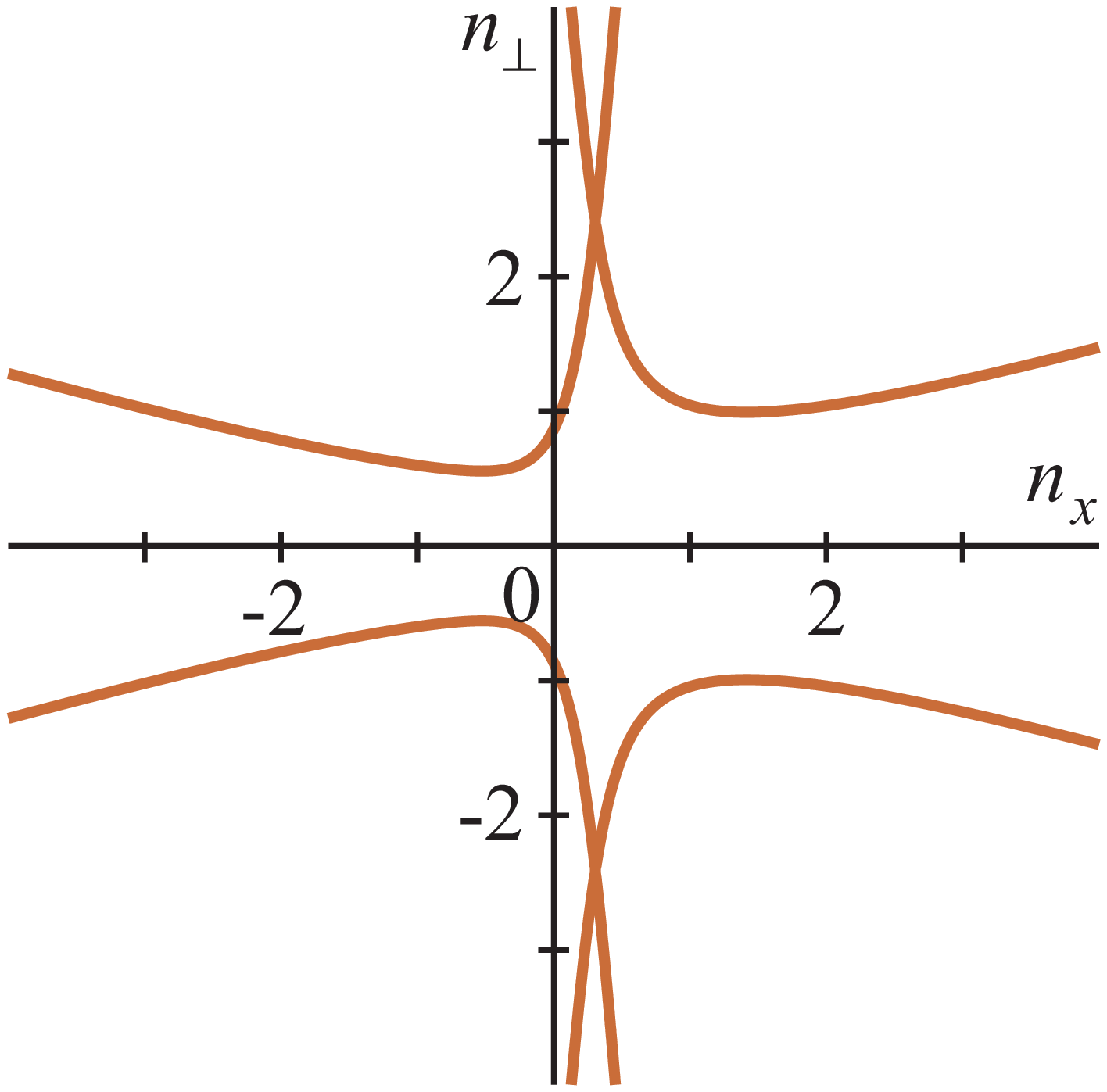}}} &
\parbox[b]{0.33\textwidth}{\centerline{\includegraphics[height=5cm]{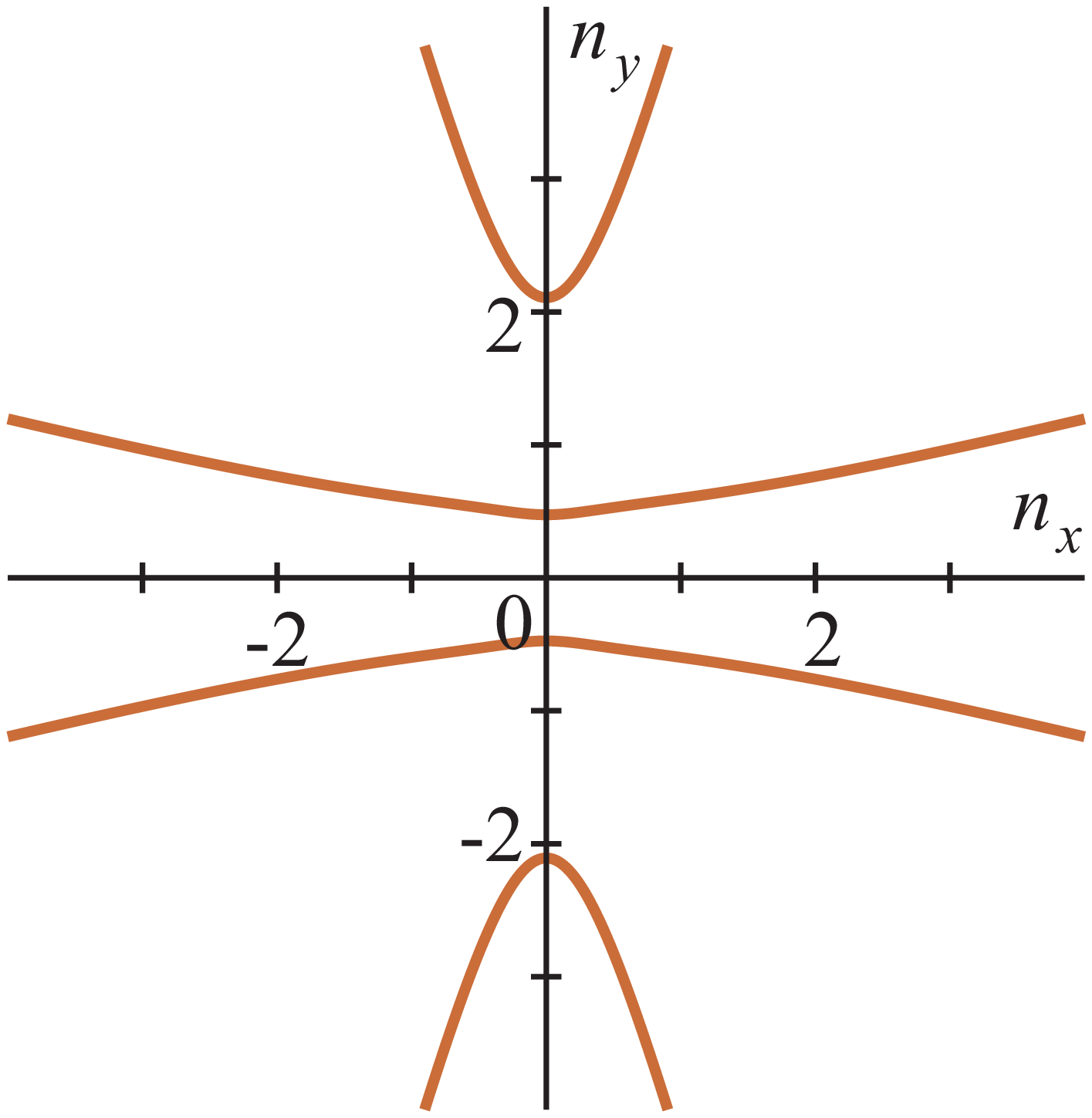}}} \cr
$(a)$ & $(b)\mathstrut $ & $(c)$ \cr}
\caption{An example of a infinite wave surface the type ${\bf I}$ ($\eps=1$, $\lambda_1=0.6+0.2i$, $\lambda_2=0.2$). The panels $(a)$ and $(b)$ correspond to cross-sections with inclination to the plane $n_z=0$ being equal to $0.43$ and $-0.43$, respectively. The panel $(c)$ describes the cross-section $n_z=0$.}\label{Iinffig}
\end{figure}

For this algebraically general type, the wave surface equation takes the sufficiently complicated form
\begin{gather}
\alpha\left(n_x^2+n_y^2+n_z^2-1\right)^2+2(n_x^2n_y^2-n_z^2)\left[\beta_1(1-\cosh 2\phi\cos 2\psi)+\beta_2(1+\cosh 2\phi\cos 2\psi)+\gamma\sinh 2\phi\sin 2\psi\right]\nonumber\\
+2(n_x^2n_z^2-n_y^2)\left[\beta_1(1+\cosh 2\phi\cos 2\psi)+\beta_2(1-\cosh 2\phi\cos 2\psi)-\gamma\sinh 2\phi\sin 2\psi\right]+4\beta_3\left(n_y^2n_z^2\cos 4\psi-n_x^2\cosh 4\phi\right)\nonumber\\
+\beta_3\left[-(n_x^2-1)^2\sinh^22\phi+(n_y^2+n_z^2)^2\sin^22\psi+2n_x(n_x^2+1)\sinh 4\phi-2n_yn_z(n_y^2-n_z^2)\sin 4\psi\right]\nonumber\\
+ 4n_x(n_y^2-n_z^2)\left[(\beta_1-\beta_2)\sinh 2\phi \cos 2\psi-\gamma \cosh 2\phi \sin 2\psi\right]\nonumber\\
- 4n_yn_z(n_x^2+1)\left[(\beta_1-\beta_2)\cosh 2\phi \sin 2\psi+\gamma \sinh 2\phi \cos 2\psi\right]\nonumber\\
+8\left[\gamma \cosh 2\phi\cos 2\psi+(\beta_1-\beta_2)\sinh 2\phi\sin 2\psi\right] n_xn_yn_z=0,
\end{gather}
where $\alpha$, $\beta_1$, $\beta_2$, $\beta_3$, and $\gamma$ are defined by (\ref{alphaX})-(\ref{gammaX}).
In order to simplify it, we will assume that $\phi=\psi=0$. Then the wave surface equation reduces to
\begin{gather}
\alpha\left(n_x^2+n_y^2+n_z^2-1\right)^2+4\beta_1(n_x^2n_z^2-n_y^2)+4\beta_2(n_x^2n_y^2-n_z^2)+
4\beta_3\left(n_y^2n_z^2-n_x^2\right)+8\gamma n_xn_yn_z=0. \label{Is}
\end{gather}
As it was demonstrated in Subsect.~\ref{IVg}, if $\eps\neq 2\Re\lambda_i$ ($i=1,2,3$) this equation defines the quartic surface with four singular points. When $\gamma=0$ all these points belong to one plane like a biaxial quasi-crystal (the Fresnelian subtype ${\bf I}_{\rm F}$). If $\gamma\neq0$ these points do not lie on a plane (the non-Fresnelian subtype ${\bf I}_{\rm NF}$). Examples of wave surfaces for both subtypes are presented in Fig.~\ref{IFfig} and Fig.~\ref{INFfig}. This surface will be infinite when at least one quantity $\beta_i/\alpha$ ($i=1,2,3$) satisfies the inequality $\beta_i/\alpha<-1$ (see, e.g., Fig.~\ref{Iinffig}); otherwise it will be finite.

As an example for the non-Fresnelian subtype ${\bf I}_{\rm NF}$, we can consider the model with $\eps=1$, $\lambda_1=-i\nu$, $\lambda_2=i\nu$, and $\lambda_3=-\lambda_1-\lambda_2=0$. From the physical point of view, it means that the dielectric permittivity tensor $\eps_i^m$ and the magnetic permeability tensor $\mu_i^m$ are equal to the Kroneker delta tensor, but there exist one non-zero component of the magneto-electric coefficients tensor. In this case, we obtain
\begin{gather}
\alpha=1,\quad \beta_1=\beta_2=\nu^2, \quad \beta_3=4\nu^2,\quad \gamma=-4\nu^3\neq0.
\end{gather}
The wave surface equation takes the form
\begin{equation}
(n_x^2+n_y^2+n_z^2-1)^2+16\nu^2(n_y^2n_z^2-n_x^2)+4\nu^2(n_x^2-1)(n_y^2+n_z^2)-32\nu^3n_xn_yn_z=0.
\end{equation}
It has four singular points which are located at
$$M_1(x_0,y_0,z_0),\quad M_2(x_0,-y_0,-z_0), \quad M_3(-x_0,y_0,-z_0), \quad M_4(-x_0,-y_0,z_0),$$
where
$$x_0=\frac{\nu}{1+2\nu^2+\sqrt{1+5\nu^2+4\nu^4}}, \quad y_0=z_0=\frac{1}{\sqrt{1+2\nu^2+\sqrt{1+5\nu^2+4\nu^4}}}.$$

For the specific case, when $\alpha=0$ and, say, $\beta_3=0$, the wave surface will always be infinite, because it contains the straight line $n_y=n_z=0$. Any cross-section through this line may contain, in addition, not more than two perpendicular straight lines.

\section{Conclusion}\label{VI}

In this paper, we have constructed a classification of dispersion relations for the non-minimal Einstein-Maxwell model with the trace-free susceptibility tensor, i.e., when the linear response tensor $C^{ikmn}$ takes the form
$$C^{ikmn}=\frac{\eps}{2}\left(g^{im}g^{kn}-g^{in}g^{km}\right)+q_3W^{ikmn},\quad W^{ikmn}g_{im}=0.$$
Since the tensor $q_3W^{ikmn}$ has the same algebraic properties as the Weyl tensor, our classification is based on    the Petrov type distribution scheme.

On the other hand, the dispersion relation can be considered as a homogeneous equation of the fourth order with respect to the wave vector components $K_m$, which determines a quartic surface in the three-dimensional real projective space $\mathbb{R}P^3$. Therefore we could classify the dispersion relations according to the number of singular points for such surfaces. From the physical point of view, the singularities relate to absence of the birefringence phenomenon  along these directions. Apart from a few specific bizarre cases (see details in the corresponding subsections of Sect.~\ref{IV}), the first and the second classification schemes appear to be interrelated:
\begin{table}[h!]
\begin{tabular}{|c|c|}
  \hline
  Petrov type & Number of singular points \\
  \hline
  ${\bf O}$ & 0 \\
  ${\bf N}$ & 1 \\
  ${\bf III}$, ${\bf D}$ & 2 \\
  ${\bf II}$ & 3 \\
  ${\bf I}$ & 4 \\
  \hline
\end{tabular}
\end{table}

As it was shown in Sect.~\ref{IV}, only for the types ${\bf N}$ and ${\bf D}$ the dispersion relation splits into two second-order equations and therefore only these types admit application of the optical (or effective) metrics approach to describe trajectories of photons. Expressions for the optical metrics in these two cases are presented in (\ref{NAB}) and (\ref{DAB}).

In Sect.~\ref{V}, for each Petrov type we have presented equations and plots of corresponding wave surfaces. As it was demonstrated, these wave surfaces can be both finite or infinite in dependence of relationships between the frame rate parameter $\phi$, the invariant scalars $\Psi_i$, and the trace parameter $\eps$. Furthermore, the Petrov type $\bf I$ should be divided into two subtypes depending on a singular points distribution of the wave surface:\\
\noindent {\it (i)} the Fresnelian subtype ${\bf I}_{\rm F}$ --- singular points lie on one plane, this subtype corresponds to a biaxial quasi-medium. For instance, this subtype arises when the magneto-electric coefficients tensor ${\nu_m}^i$ vanishes;\\
\noindent {\it (ii)} the non-Fresnelian subtype ${\bf I}_{\rm NF}$ --- singular points do not lie on one plane; for instance, this subtype arises when the dielectric permittivity tensor $\eps_m^i$, the magnetic impermeability tensor $(\mu^{-1})_m^i$, and the magneto-electric coefficients tensor ${\nu_m}^i$ take the form
$$\eps_m^i=(\mu^{-1})^i_m=\delta^i_m, \quad \nu_m^i=\nu\left(\delta_1^i\delta^1_m-\delta_2^i\delta^2_m\right).$$
Thus, in the framework of the non-minimal Einstein-Maxwell model with the trace-free susceptibility tensor, we can divide the set of dispersion relations into {\it seven} basic sorts (${\bf I}_{\rm F}$, ${\bf I}_{\rm NF}$, $\bf II$, $\bf III$, $\bf D$, $\bf N$, and $\bf O$) according to number and position of their singular points.

The Petrov classification scheme appears to be very productive and perspective to arrange dispersion relations and wave surface types. However, it does not involve a lot of interesting cases \cite{BZ2008,Itin2010,Favaro4,Hehl4,Itin2015,Favaro3}, e.g., the case for which the wave surface has 16 {\it real} singular points \cite{Favaro3} (in contrast with the type $\bf I$, where the surface possesses only 4 {\it real} singular points). Therefore we are going to broaden the scope of our classification and try to apply this approach to the models with the non-vanishing tensor $S_{mn}=R_{mn}-\frac14 Rg_{mn}$ and, in principle, the non-vanishing skewon part of the linear response tensor $C^{ikmn}$.
We believe that this detailed classification will help us to find new types of dispersion relations and new types of wave surfaces and to study other geometrical optics effects: distortion, caustics etc.

\appendix

\acknowledgments
Authors are grateful to J.~P.~S.~Lemos and V.~Perlick for fruitful discussions and advices.
The work was supported by Russian Science Foundation (Project No. 16-12-10401), and, partially, by the Program of Competitive Growth of Kazan Federal University.

\end{document}